\documentclass[twocolumn,reprint,superscriptaddress,amsmath,amssymb,prb,aps]{revtex4-2}

\usepackage{graphicx}
\usepackage{bm}
\usepackage{bbm}
\usepackage{hyperref}
\usepackage{braket}
\usepackage{soul}
\usepackage{color}
\usepackage{orcidlink}
\usepackage{lipsum}
\usepackage{tikz}
\usepackage{glossaries}
\usetikzlibrary{quantikz2}

\hypersetup{
    colorlinks=true,
    linkcolor=blue,
    filecolor=magenta,      
    urlcolor=blue,
    citecolor=blue
}

\begin{document}

\newacronym{aqc}{AQC}{approximate quantum compiling}
\newacronym{mps}{MPS}{matrix product state}
\newacronym{dmrg}{DMRG}{density matrix renormalisation group}
\glsdisablehyper

\title{Quench Spectroscopy of Magnetic Excitations on a Superconducting \\Quantum Processor}

\author{D.~A.~Millar\,\orcidlink{0000-0003-3713-8997}}
\email{declan.millar@ibm.com}
\affiliation{IBM Research, Hursley, Winchester, SO21 2JN, United Kingdom}
\affiliation{School of Physics and Astronomy, University of Southampton, Southampton SO17 1BJ, UK}

\author{G.~W.~Pennington\,\orcidlink{0009-0008-9257-5878}}
\affiliation{The Hartree Centre, STFC, Sci-Tech Daresbury, Warrington WA4 4AD, United Kingdom}

\author{N.~T.~M.~Siow}
\affiliation{London Centre for Nanotechnology, University College London, Gordon St., London, WC1H 0AH, United Kingdom}

\author{S.~Brandhofer}
\affiliation{IBM Research, Ehningen, 71139, Germany}

\author{J.~Crain\,\orcidlink{0000-0001-8672-9158}}
\affiliation{IBM Research, Hursley, Winchester, SO21 2JN, United Kingdom}
\affiliation{Clarendon Laboratory, University of Oxford, Oxford OX1 3PU, UK}

\author{F.~H.~L.~Essler\,\orcidlink{0000-0002-1127-5830}}
\affiliation{Clarendon Laboratory, University of Oxford, Oxford OX1 3PU, UK}

\author{A.~G.~Green\,\orcidlink{0000-0002-3923-5291}}
\affiliation{London Centre for Nanotechnology, University College London, Gordon St., London, WC1H 0AH, United Kingdom}

\author{S.~J.~Thomson\,\orcidlink{0000-0001-9065-9842}}
\email{steven.thomson@ed.ac.uk}
\affiliation{SUPA, School of Physics and Astronomy, University of Edinburgh, Peter Guthrie Tait Road, Edinburgh EH9 3FD, UK}

\date{\today}

\begin{abstract}
The elementary excitation spectrum of a many-body quantum system encodes many key properties, including phenomena as diverse as transport, thermalisation and ground state structure. Excitation spectra of strongly correlated systems are typically encoded in dynamical structure factors, which are demanding to measure experimentally and challenging to compute classically. Here we use \emph{quench spectroscopy} on a superconducting quantum processor to extract excitation spectra of spin chains of $L=101$ spins. By tailoring the combination of quench protocol and observable, we selectively access distinct excitation sectors across several phases of the spin-$1/2$ XXZ chain, resolving free magnons, multi-magnon bound states, and two-spinon continua.
Notably, we demonstrate that the protocol does not rely on ground state preparation: in the classically challenging gapless regime, we extract spectra directly from the quench dynamics of easily prepared product states, a procedure that is natural and straightforward on quantum hardware.
Our work establishes quench spectroscopy as a fast and flexible probe of many-body excitation spectra on digital quantum hardware, introduces a novel quench protocol that does not require costly state preparation routines, and provides a scalable route towards regimes where classical simulation may become intractable.
\end{abstract}

\maketitle

\begin{figure*}[!ht]
    \centering
    \includegraphics[width=\linewidth]{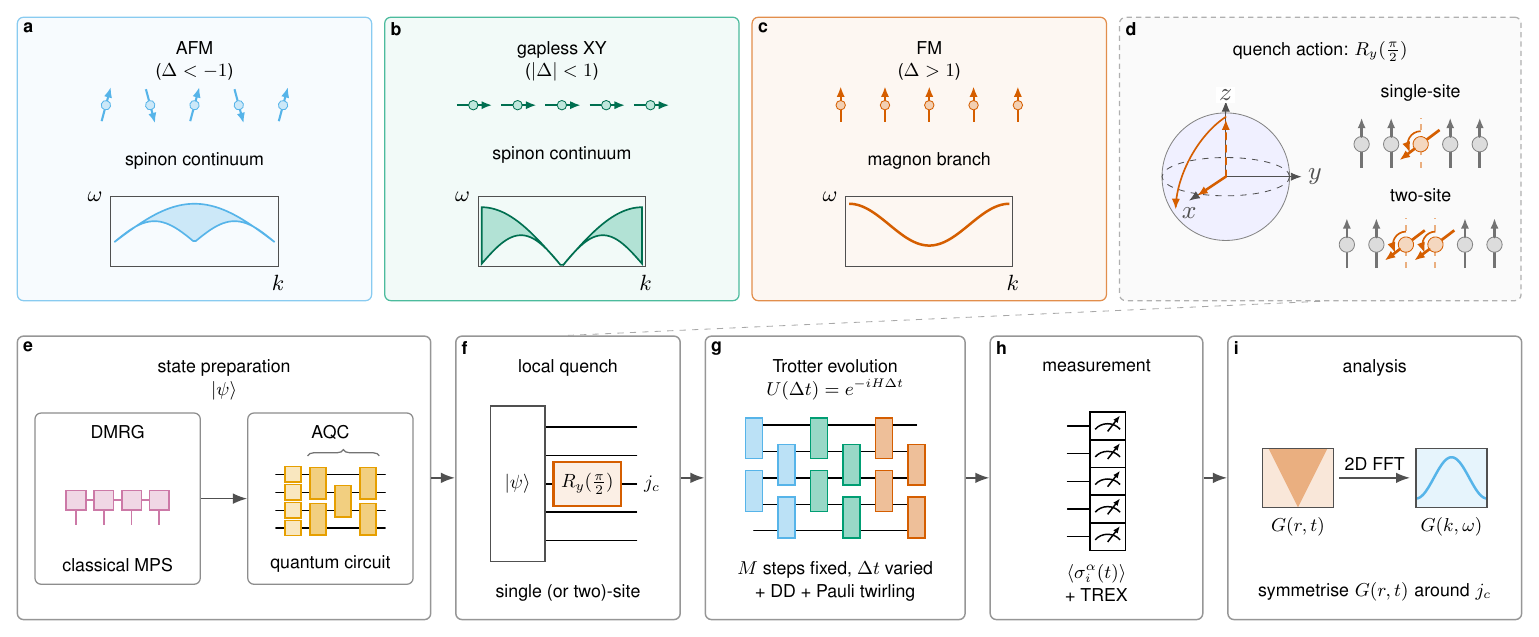}
    \caption{Local quench spectroscopy on a digital quantum computer. (a-c)~The three phases of the spin-$\tfrac{1}{2}$ XXZ chain considered in this work (AFM, gapless XY, FM) differ qualitatively in their ground state structure, the cost of preparing that state on hardware, and the expected spectroscopic signature. (d)~The action of the local quench in the FM phase. (e)~Hardware realisation: in the FM and XY phase the pre-quench state is a product state; in the AFM phase we compute a classical ground state using density matrix renormalisation group (DMRG), then compile the resulting matrix product state (MPS) to a shallow brickwork circuit using approximate quantum compiling (AQC). (f)~The state preparation circuit is followed by a local $R^{j_c}_y(\pi/2)$ quench applied to one or more sites. (g)~Trotterised time evolution with a fixed number of steps $M$ but variable step size $\Delta t$, and (h)~measurement of $\sigma^\alpha$ across the chain. (i) Classically symmetrise $G(r,t)$ around the quench site and apply a 2D Fourier transform to give the quench spectral function $G(k,\omega)$. Hardware-level error mitigation is applied: dynamical decoupling (DD) and Pauli twirling during the evolution, and twirled readout error extinction (TREX) at the readout step.}
    \label{fig:schematic}
\end{figure*}

\section*{Introduction}

The investigation of many-body quantum matter is one of the most promising applications of near-term quantum simulators, which offer an unprecedented ability to engineer complex quantum states at a microscopic level. Large-scale analog simulators can build precise models of quantum matter a single atom at a time, while current-generation digital simulators are smaller, but offer the flexibility to investigate a wider range of models. Many of the physical systems we are most interested in are classically accessible only in restricted regimes, such as weak entanglement or perturbative interactions. The study of strongly correlated quantum materials therefore provides a natural setting in which quantum simulation can directly access regimes beyond the reach of classical methods, where qualitatively new phenomena may emerge.

A key quantity of interest in condensed matter physics and materials science is the \emph{elementary excitation spectrum}, which encodes a wealth of information about the transport properties of a quantum system, and therefore offers a valuable window into the physical properties of a given material. Excitation spectra can be read off from spectral functions or dynamical structure factors, which are typically obtained by Fourier transforming unequal-time two-point correlation functions~\cite{Sturm93}. Experimentally, these are measurable using a variety of pump-probe techniques including angle-resolved photoemission spectroscopy, X-ray Raman scattering, Bragg spectroscopy and inelastic neutron or muon scattering~\cite{Jia+14, Gao+24}.
For general Hamiltonians, computing the dynamics of unequal-time correlators (and thus computing spectral functions) is bounded-error quantum polynomial time (BQP)-hard in many practical cases~\cite{Baez+20}. This renders exact classical solutions intractable for large systems---although powerful approximate methods exist in both one and two dimensions, including tensor networks~\cite{Schollwoeck+11,Bridgeman+17,Bera+17,Tindall+24,Tindall+25}, neural quantum states~\cite{Carleo+17}, and continuous unitary transforms~\cite{Thomson+24}.
This motivates the use of quantum simulators as tools to probe excitation spectra, as quantum hardware is in principle able to realise exact non-equilibrium evolution natively, without approximation.
Even on quantum hardware, however, such spectral probes remain operationally challenging to measure. Reconstructing unequal-time correlation functions either requires carefully designed experimental setups~\cite{Landig+15} or the use of further approximations to reconstruct them from a series of measurements via linear response theory~\cite{Knap+13,Baez+20,Lee+26}. Here we employ a different type of dynamical spectroscopic probe that is local in space and time, does not require any hardware-specific protocol or approximation, and can be applied to quantum systems of any dimensionality or geometry.

The core technique we will use in this work is \emph{quench spectroscopy}~\cite{Gritsev+07, Menu+18, villa2019unraveling, villa2020local,Villa21}. 
This method builds on the intuitive idea that the non-equilibrium dynamics of a quantum system encode important information about the motion of elementary excitations~\cite{Frerot+18,Cevolani+18,Despres+19}. The main object of interest in quench spectroscopy is the \emph{quench spectral function} (QSF), which was introduced in Refs.~\cite{villa2019unraveling,villa2020local,Villa21}, and is given by the Fourier transform of a single-time observable. The conventional procedure, shown schematically in Fig.~\ref{fig:schematic}, is to prepare the system in its ground state, perform a quench that generates the desired type of excitation, and then follow the out-of-equilibrium dynamics resulting from the quench. For an appropriately chosen combination of quench and observable, a straightforward Fourier transform then leads directly to the excitation spectrum. The quench may be \emph{global} (involving changes to the underlying Hamiltonian), or \emph{local} (such as flipping or rotating a single spin).

In the case of local quenches~\footnote{More generally, this form holds for any initial state which breaks translation invariance.}, which we shall focus on here, the QSF is defined as the space-time Fourier transform of a local observable $G({\bf r},t) = \langle O({\bf r}, t) \rangle = \textrm{Tr}[\rho_0 O({\bf r},t)]$, given by:
\begin{align}
G({\bf k},\omega) &= \int dr\, dt \, e^{-i({\bf k} \cdot {\bf r} - \omega t)} G({\bf r},t) \nonumber \\
&= (2 \pi)^2 \sum_{n, n'} \rho_0^{n,n'} \braket{n' | O| n} \nonumber \\
& \quad \quad \times \delta({\bf P}_{n} - {\bf P}_{n'} - {\bf k}) \delta(E_n - E_{n'} - \omega) 
\label{eq:qsf}
\end{align}
where $\rho_0$ represents the initial density matrix immediately following the quench, and $n,n'$ label eigenstates. This quantity exhibits peaks at energies $\omega = E_{n} - E_{n'}$ and momenta ${\bf k} = {\bf P}_{n} - {\bf P}_{n'}$. In order for this to be non-zero, the following conditions must be satisfied~\cite{villa2020local,Villa21}:
\begin{align}
\langle n' | O | n \rangle & \neq 0, \label{eq:coupling_condition1}\\
\rho_0^{n, n'} = \braket{n | \psi_0} \braket{\psi_0 | n'} &\neq 0. \label{eq:coupling_condition2}
\end{align}
By appropriate choices of the initial state, quench, and observable, the QSF is able to resolve elementary excitations. 
The key ingredients are that the quench should put the system into a coherent superposition of the target manifolds---one of which may be the ground state, though this is not a requirement---and the observable should couple these manifolds, allowing transitions between them to be probed. Compared with other spectroscopic protocols~\cite{Knap+13,Baez+20,Lee+26}, quench spectroscopy requires no linear response assumption and no reconstruction of Green's functions from multiple observables, instead obtaining the spectrum directly from the evolution of a single observable, local in both space and time.

Quench spectroscopy was originally developed in the context of ultracold atomic gases~\cite{villa2019unraveling, villa2020local,Villa21}. It has since been applied to disordered systems~\cite{Villa+21a, Villa+21b}, lattice gauge theories~\cite{Chanda+24}, fermionic and dissipative systems~\cite{Bocini+25,Despres25}, and has recently been realised in both analog and small-scale digital quantum simulations~\cite{Chen+25, Sun+25, VilchezEstevez+25}. Digital quantum processors are particularly well suited for this approach as they avoid the need for harmonic confining potentials---which can subtly modify the underlying physics on analog platforms~\cite{Yu+25}---allowing more direct implementation of the Hamiltonian and quench protocol.

Here we demonstrate the large-scale use of local quench spectroscopy on a superconducting quantum processor to accurately extract excitation spectra of length $L=101$ quantum spin chains. Our motivation in this work is twofold. First, we demonstrate that spectral features of large, complex quantum systems can be directly measured on current hardware using quench spectroscopic techniques, with no signal reconstruction or extrapolation required. Across three phases of the one-dimensional XXZ model, we access a wide range of spectral properties: magnons and magnon bound states in the ferromagnetic phase, two-spinon continua in the antiferromagnetic phase, and magnon-like excitations in the gapless phase. The success of this technique (benchmarked against exact analytical solutions) validates the performance of the quantum hardware, confirming that it is faithfully encoding the desired microscopic model, despite the presence of noise and experimental imperfections. Secondly, we also demonstrate that detailed knowledge of the ground state of the system is not required for spectroscopy, and we propose a general symmetry-based heuristic for finding suitable initial states that can be prepared at low circuit depth on near-term quantum computers.

\section*{Results}

\subsection*{Model} 

We consider the spin-$\tfrac{1}{2}$ XXZ chain with open boundary conditions:
\begin{equation}
H = -J\sum_{i=1}^{N-1} \left(S^x_i S^x_{i+1}
+ S^y_i S^y_{i+1}
+ \Delta S^z_i S^z_{i+1} \right),
\label{eq:hamiltonian}
\end{equation}
where ${S}^\alpha_i$ ($\alpha \in [x,y,z]$) is the spin operator, related to the Pauli matrices $\sigma^\alpha_i$ via ${S}^\alpha_i = \tfrac{1}{2}{\sigma}^\alpha_i$. This model exhibits two quantum phase transitions at $\Delta = \pm 1$. Throughout this work, we fix $J=1$ as our unit of energy and vary $\Delta$ to explore the ferromagnetic ($\Delta > 1$), gapless ($|\Delta| < 1$), and antiferromagnetic ($\Delta < -1$) regimes. This Hamiltonian is integrable~\cite{korepin1993quantum,takahashi1999thermodynamics,gaudin2014bethe} and its elementary excitation spectrum is exactly known across all three phases, making it an ideal non-trivial benchmark for our purposes. As a consequence of its integrability, it supports stable excitations not only in the ground state sector, but at arbitrary energy densities. The properties of these excitations depend on the details of the finite energy density macrostate the system relaxes to~\cite{bonnes2014light,essler2016quench}. However, at very low energy densities above the ground state the dispersion relations of these excitations will be close to the ones of excitations above the ground state. 

\begin{figure*}
    \centering
    \includegraphics[width=\linewidth]{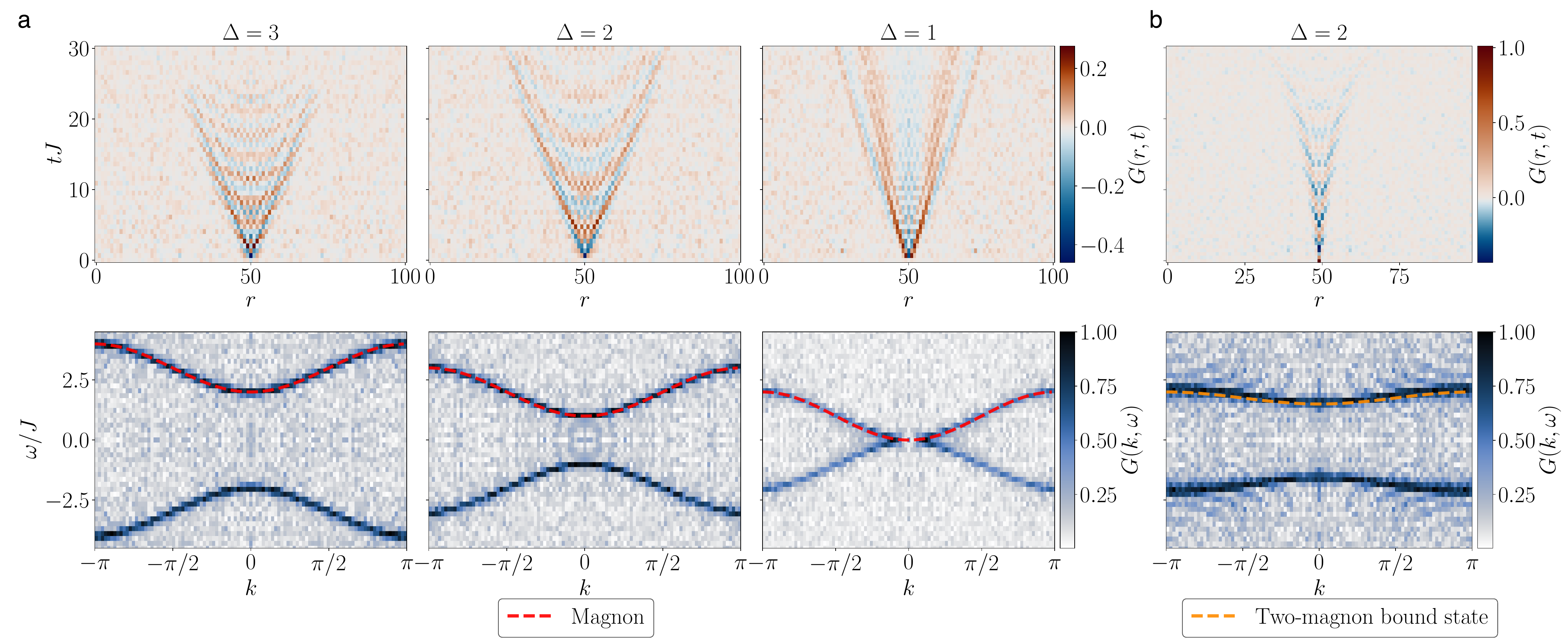}
    \caption{Local quench spectroscopy in the ferromagnetic phase of the XXZ chain. (a) Single-site $R_y(\tfrac{\pi}{2})$ quenches for $\Delta=3,2,1$, respectively, showing the measured real-space dynamics $G(r,t)=2\langle S^y(r,t)\rangle$ (top) and the corresponding quench spectral functions $G(k,\omega)$ (bottom). The red lines indicate the exact single-magnon dispersion from Bethe ansatz, Eq.~\ref{eq.fm_dispersion}. 
    (b) Two-site $R^{j}_y(\tfrac{\pi}{2})R^{j+1}_y(\tfrac{\pi}{2})$ quench for $\Delta=2$ with observable Eq.~\ref{eq:two_magnon_bound_dispersion}. The orange line in the lower panel indicates the bound state dispersion from the Bethe ansatz, Eq.~\ref{eq:two_magnon_bound_dispersion}.}
     \label{fig:fm}
\end{figure*}

\subsection*{Ferromagnetic Phase: Magnons}

We first demonstrate the principle of local quench spectroscopy in the ferromagnetic phase ($\Delta > 1$). In this regime, the ground state is a fully polarised product state that can be prepared exactly using only single-qubit gates, without requiring variational methods or approximate circuit compilation. The elementary excitations in this regime are magnons, which are essentially linear combinations of single spin flips. In the ferromagnetic XXZ chain, the single-magnon dispersion is given by:
\begin{equation}
E(k) = J(\Delta - \cos k),
\label{eq.fm_dispersion}
\end{equation}
which follows from the Bethe ansatz solution~\cite{takahashi1999thermodynamics,villa2020local}. The relatively simple structure of both the state and the elementary excitations serves as a good starting point to illustrate the method.
In this regime, the local quench takes the form of a single-site $R_y(\pi/2)$ rotation applied to the central site $j_c$, i.e. $\ket{\psi(0)} = R^{j_c}_y(\pi/2) \ket{\uparrow \uparrow ...}$~\cite{villa2020local}. Following this quench, we measure the local observable $2 S^y(r,t) = \sigma^y(r,t)$ across the chain. Note that we make a $R_y(\pi/2)$ rotation, as a full spin flip (i.e. a $R_y(\pi)$ rotation) produces a state orthogonal to the ground state, such that the ground-to-excited-state coherences vanish and the desired spectroscopic signal is suppressed~\cite{villa2020local}. This reflects the coupling condition discussed in Eq.~\ref{eq:coupling_condition}. The resulting space--time signal $G(r,t)=\langle \sigma^y(r,t)\rangle$ is symmetrised around the quench site and Fourier transformed to obtain the quench spectral function (QSF) $G(k,\omega)$.

Figure~\ref{fig:fm}(a) shows the resulting real-space dynamics $G(r,t)$ and the corresponding quench spectral function (QSF) $G(k,\omega)$ for $\Delta \in [3,2,1]$. All results presented in the main text are directly obtained from experiments performed on the \texttt{ibm\_boston} quantum processor. We do not employ any additional processing steps other than dynamical decoupling, Pauli twirling and twirled readout error extinction (TREX), all of which are standard error mitigation techniques integrated into \texttt{Qiskit}~\cite{qiskit2024}. Notably, we do not make use of Zero Noise Extrapolation in our results. The surprising robustness of the Fourier-transformed signal to device noise is likely due to the lack of space-time correlations resulting in noise with a flat Fourier spectrum~\cite{Foldager+23,Dalzell+24,Sun+25}. For details of the circuit implementation, see Ref.~\cite{SM}, which also contains information for all experiments discussed in later sections.

In real space, the dynamics exhibit a clear light-cone structure, indicating the propagation of excitations generated by the local perturbation~\cite{Ganahl2012}. While the overall extent of the light cone remains independent of $\Delta$ as the maximal group velocity does not change, its internal oscillatory structure changes noticeably as $\Delta$ is varied. The QSF also evolves distinctly with $\Delta$. As $\Delta \to 1$, the excitation gap closes continuously, signalling the approach to the critical point between the gapped ferromagnetic phase and the gapless XY phase (classically studied with quench spectroscopy in Ref.~\cite{villa2020local}). This is directly visible in $G(k,\omega)$, where the minimum excitation energy approaches $\omega=0$ at $k=0$. For all values of $\Delta$, the quench spectral function exhibits a sharply defined dispersive mode.
The extracted dispersion is in quantitative agreement with this prediction, with the experimentally obtained spectral weight lying directly on top of the theoretical curve, thereby identifying the observed mode as the single-magnon excitation.
The observed light-cone structure is also consistent with the propagation of single-magnon quasiparticles. The leading edge of the light cone is set by the maximal group velocity $v_{\mathrm{max}}=\max_k \partial_k E(k)=|J|$, which is independent of $\Delta$. The internal oscillatory structure reflects interference between modes with different group velocities, determined by the $\Delta$-dependent dispersion $E(k)$, together with their momentum-dependent excitation weight set by the specifics of the quench~\cite{villa2020local}.

This demonstrates that near-term quantum computers can accurately capture collective excitations in large many-body quantum systems, and remarkably that the accurate measurement of the spectrum only requires lightweight quantum error mitigation techniques whose sampling overhead remains independent of system size. Having shown that the fundamentals of the method can be successfully executed on quantum hardware, we are now in a position to move on to more exotic phenomena.

\subsection*{Ferromagnetic Phase: Magnon Bound States}

In the ferromagnetic regime, elementary excitations include multi-magnon bound states. In general, an $m$-magnon eigenstate can be written in the form~\cite{Ganahl2012}:
\begin{equation}
|\{k_j\}\rangle
=
\sum_{x_1<\cdots<x_m}
\Psi(\{k_j\}|\{x_l\})
\prod_{n=1}^{m} S^+_{x_n}\,|0\rangle,
\label{eq.multi-magnon}
\end{equation}
where $|0\rangle$ denotes the fully polarized ferromagnetic ground state and $\Psi$ is a Bethe-ansatz wavefunction.
Here $x_l$ label the lattice positions of the $m$ flipped spins, while the
Bethe quasi-momenta $k_j$ parametrise the eigenstate. For real $k_j$ the
state describes scattering magnons, whereas complex solutions correspond to string bound states for which the wavefunction is exponentially localized in the relative separations $|x_l-x_{l'}|$.

In the two-magnon sector, the wavefunction factorizes into centre-of-mass and relative coordinates, with a total momentum $k$ and a relative-coordinate wavefunction $\Psi(r)$ that is exponentially localized for a bound state. In the strong-binding limit $\Delta \gg 1$, this localization is strongest at nearest-neighbour separation, so that the state is dominated by configurations of the form $S^+_j S^+_{j+1}|0\rangle$~\cite{Ganahl2012}.
In the thermodynamic limit, the two-magnon bound-state dispersion is given by~\cite{takahashi1999thermodynamics}:
\begin{equation}
    E_{2}(k)
    =
    J\,
    \frac{\sinh \eta}{\sinh 2\eta}
    \left(\cosh 2\eta-\cos k\right),
\label{eq:two_magnon_bound_dispersion}
\end{equation}
where $\eta=\operatorname{arccosh}\Delta$.
Similar bound states have previously been reconstructed from $n$-body correlation functions in a quantum simulation experiment of the periodically driven XXZ chain~\cite{Morvan2022}, where it was demonstrated (and later theoretically confirmed~\cite{Hudomal+24}) that they are robust to integrability-breaking perturbations. Motivated by the above analysis, here we demonstrate that bound states in the XXZ chain can be directly observed using a two-site local quench combined with a two-site observable. Concretely, we apply a quench $L = R_y^{j_c}(\tfrac{\pi}{2})\, R_y^{j_c+1}(\tfrac{\pi}{2})$ to the central pair of sites and measure the two-site operator:
\begin{align}
 O(j)
&= 4\left(S_j^+S_{j+1}^+ + S_j^-S_{j+1}^-\right) 
= 2\left(S_j^xS_{j+1}^x - S_j^yS_{j+1}^y\right),
\label{eq:bound_state_observable}
\end{align}
which is the Hermitian form of the bound-state creation and annihilation operator, and is therefore naturally sensitive to the bound-state sector, while not coupling to single magnon excitations.

The two-site quench $L = R_y^{j_c}(\tfrac{\pi}{2}) R_y^{j_c+1}(\tfrac{\pi}{2})$ generates products of single-spin rotations which produce adjacent spin-flip pairs $\propto S^y_{j_c}S^y_{j_c+1}$. The resulting post-quench state contains contributions from configurations with both individual and neighbouring flipped spins, yielding finite overlap with both two-magnon scattering states and bound states. The two-site observable, Eq.~\ref{eq:bound_state_observable}, then selectively projects onto this sector, suppressing single-magnon contributions, since $\langle \text{single-magnon} | O | 0 \rangle = 0$ for operators that create spin-flip pairs.

Figure~\ref{fig:fm}(b) shows the resulting real-space dynamics $G_{\mathrm{b}}(r,t)$ and the corresponding quench spectral function $G_{\mathrm{b}}(k,\omega)$. Strikingly, the leading edge of the bound-state light-cone bends outwards, which is an artefact of the finite and variable Trotter step size increasing at longer evolution times. A thorough analysis of this effect for both single and bound magnons is shown in Ref.~\cite{SM}. The QSF exhibits a distinct branch in excellent quantitative agreement with the predicted bound-state branch in Eq.~\ref{eq:two_magnon_bound_dispersion} across the Brillouin zone, further demonstrating the robustness of this bound state to noise and other integrability-breaking effects~\cite{Hudomal+24}. Having shown that spectra can be recovered starting from eigenstates that are easy to prepare, now we move on a a scenario where the initial state preparation is more demanding.

\begin{figure}[!tb]
    \centering
    \includegraphics[width=\linewidth]{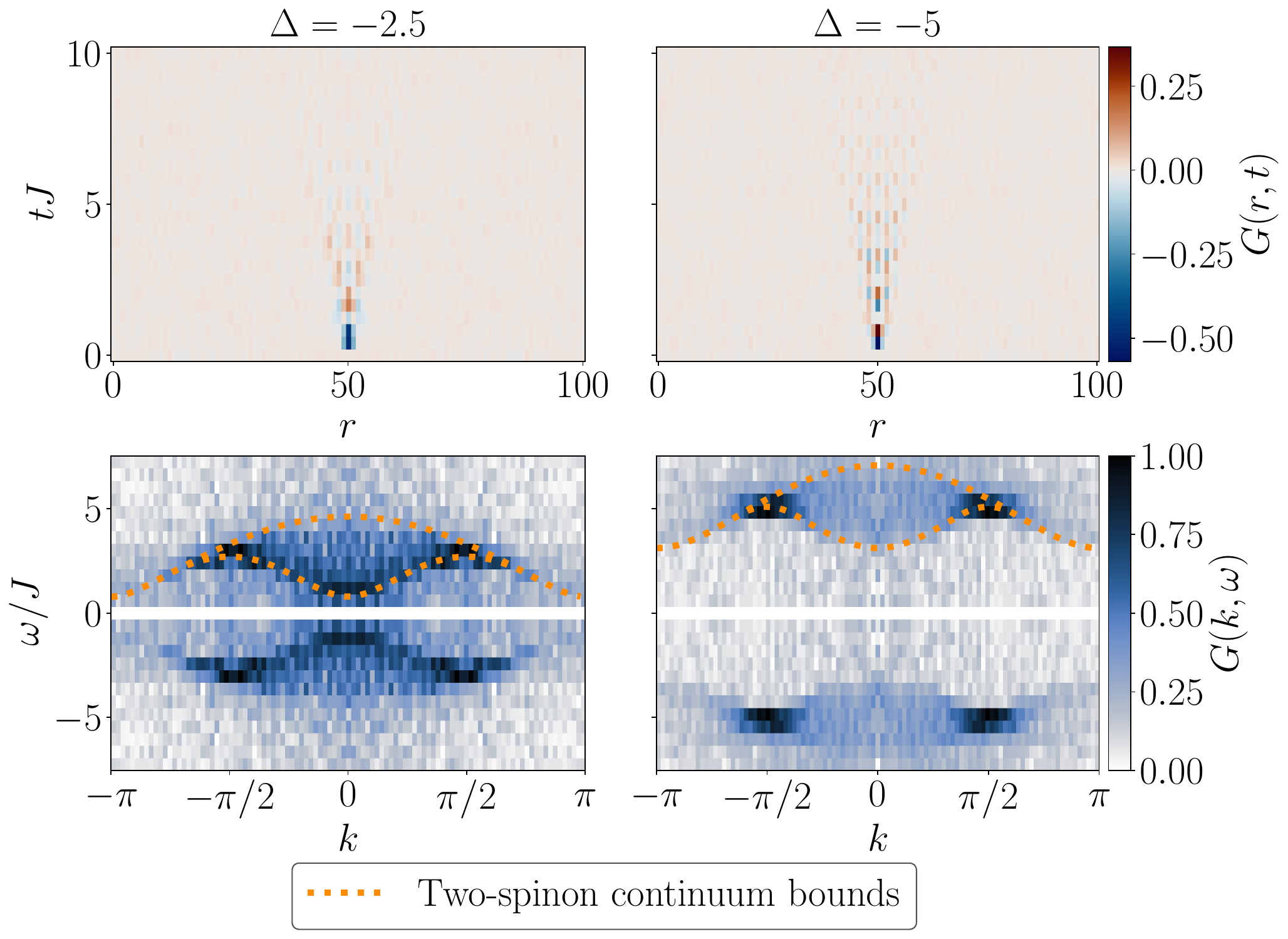}
    \caption{Local quench spectroscopy in the antiferromagnetic phase with a $R_y(\pi/2)$ single-site quench showing real-space dynamics $G(r,t)=2\langle S_y(r,t) \rangle$ (top) and quench spectral functions $G(k,\omega)$ (bottom) for $\Delta=-2.5, 5.0$. In this regime, the QSF displays a broad two-spinon continuum (Eq.~\ref{eq:two_spinon_energy}) bounded by the exact Bethe-ansatz thresholds (orange lines). }
    \label{fig:afm}
\end{figure}

\subsection*{Antiferromagnetic Phase: Spinon Continuua}

In contrast to the ferromagnetic phase, where the ground state is a fully polarized product state that can be prepared exactly using only single-qubit gates, the antiferromagnetic (AFM) regime ($\Delta < -1$) presents a qualitatively different starting point. In the Ising limit $\Delta \to -\infty$, the system becomes twofold degenerate and approaches a superposition of the two N\'eel product states $\ket{\uparrow\downarrow\uparrow\downarrow\cdots}$ and $\ket{\downarrow\uparrow\downarrow\uparrow\cdots}$. For finite $\Delta < -1$, transverse spin-exchange terms induce local spin flips that generate superpositions of configurations with domain-wall excitations on top of the N\'eel backgrounds, resulting in an entangled ground state with short-range correlations.

Because the gapped antiferromagnetic phase has a finite correlation length, the ground state admits an efficient matrix product state (MPS) representation with moderate bond dimension. We obtain this state classically using density matrix renormalization group (DMRG). In principle, MPSs can be mapped directly to quantum circuits using sequential or compressed MPS-to-circuit constructions~\cite{schon2005sequential,schon2007sequential,ran2020encoding,lin2021real,mpstocircuit2025}. In practice, however, these exact mappings can produce deep circuit structures that are less well suited to near-term hardware than shallow variational ansatzes. We therefore use the DMRG state as a classical target for approximate quantum compilation and optimize a shallow nearest-neighbour brickwork circuit to maximise the fidelity~\cite{robertson2025approximate,qiskit-addon-aqc-tensor}.

For $\Delta=-2.5$, we prepare the antiferromagnetic ground state on quantum hardware by compiling a DMRG solution for a system of $L=101$ spins into a four-layer nearest-neighbour brickwall circuit with total CNOT depth 24. The DMRG calculation was performed allowing the bond dimension to grow up to $\chi=64$, ensuring a highly accurate approximation of the target state, though in practice, much smaller bond dimensions already yield near-converged states (see Ref.~\cite{SM} for details). The resulting compiled circuit achieves a state fidelity of $F=0.989$ with respect to this target MPS, though it is not guaranteed to preserve translation invariance or inversion symmetry. For the more strongly gapped case $\Delta=-5$, the ground state is substantially less entangled allowing compilation to a brickwork circuit with CNOT depth 12 and a fidelity $F=0.999$. 

These results suggests that, for this system size and parameter regime, the ground state can be well approximated by a low-depth circuit representation, consistent with recent results demonstrating that nontrivial gapped one-dimensional ground states can be compiled into shallow brickwork circuits with high fidelity on systems of comparable size~\cite{pennington2026preparing}. See Methods for further details of the ground state preparation procedure.

In the $\Delta < -1$ regime, the elementary excitations are spinons, which can be understood as mobile domain walls separating the two N\'eel ordering patterns. Defining $|\Delta|=\cosh\gamma$, the single-spinon dispersion is given by:
\begin{equation}
E(q) = \frac{2\mathbf{K}(u)}{\pi}\,J\sinh\gamma\, \sqrt{1 - u^{2}\cos^{2} q},
\label{eq:single_spinon_dispersion}
\end{equation}
where $\mathbf{K}(u)$ is the complete elliptic integral of the first kind, and the elliptic modulus $u$ is determined by $\mathbf{K}'(u)/\mathbf{K}(u) = \gamma/\pi$, with $\mathbf{K}'(u) \equiv \mathbf{K}(\sqrt{1-u^{2}})$~\cite{takahashi1999thermodynamics}.

Although the elementary excitations are individual spinons, local spin operators do not couple the ground state directly to isolated spinons, but instead predominantly excite pairs of spinons. This reflects the fractionalisation of local integer spin excitations into mobile spin-$\tfrac12$ quasiparticles. As a result, the expected spectroscopic signature is not a sharp quasiparticle branch, but a broad two-spinon continuum~\cite{takahashi1999thermodynamics}, as seen in inelastic neutron scattering data on real XXZ AFM materials~\cite{satija1980neutron,yoshizawa1981dynamical,goff1995exchange,Bera+17}.

For a two-spinon state with total momentum $Q$, the individual spinon momenta satisfy $q_{1}+q_{2} = Q + \pi \pmod{2\pi}$, where the additional $\pi$ shift reflects the doubled unit cell of the antiferromagnetic ground state. The total excitation energy is:
\begin{equation}
E_{2\mathrm{s}}(Q,q) = E(q) + E(Q + \pi - q).
\label{eq:two_spinon_energy}
\end{equation}
The lower continuum boundary is obtained by minimizing $E_{2\mathrm{s}}(Q,q)$ over $q$. The upper boundary is obtained from the symmetric momentum partition $q_1=q_2=(Q+\pi)/2$, giving $E_{\mathrm{U}}(Q)=2E\!\left(\frac{Q+\pi}{2}\right)$.

We probe these excitations experimentally by preparing the compiled AFM ground states, applying a local $R_y(\pi/2)$ to the centre of the chain and measuring the resulting dynamics of $\langle \sigma^y(r,t)\rangle$, which satisfies the coupling conditions of Eqs~\ref{eq:coupling_condition1} and \ref{eq:coupling_condition2}. The system is then evolved under the XXZ Hamiltonian up to a total time $tJ=10$ using a fixed-depth second-order Trotter decomposition with 15 Trotter steps, yielding a time-evolution circuit with CNOT depth 93. Combined with the approximately compiled initial states for $\Delta=-2.5$ and $5.0$, this yields circuits with 117 and 105 CNOT depth, respectively. The dynamics are sampled at 25 time slices, from which we reconstruct the space--time signal and corresponding quench spectral function via Fourier transformation.

Figure~\ref{fig:afm} shows the resulting real-space dynamics $G(r,t)$ and the corresponding quench spectral function $G(k,\omega)$ for $\Delta=-2.5$ and $\Delta=-5$. In real space, the dynamics exhibit a clear light-cone structure with internal interference patterns reflecting the broad distribution of spinon group velocities. In contrast to the ferromagnetic case, the spectral response does not form a sharply defined branch; instead, the spectral response forms a broad continuum. The measured spectral weight is confined within the exact two-spinon bounds and is strongly concentrated near the lower edge of the continuum.
This behaviour is in quantitative agreement with the expected two-spinon continuum and is characteristic of fractionalized spinon excitations in one dimension. The antiferromagnetic data therefore show that local quench spectroscopy on quantum hardware can resolve not only sharp quasiparticle dispersions, but also broad many-body continua associated with fractionalisation.

\subsection*{XY Regime: Magnon-like Excitations}

Finally, we turn to the critical XY regime, $|\Delta| < 1$, where the ground state is $U(1)$ symmetric, supports gapless excitations and exhibits Luttinger liquid behaviour~\cite{takahashi1999thermodynamics}. The precise excitation structure in this parameter regime is complex, with multiple distinct types of excitation possible depending on the sign of $\Delta$~\cite{takahashi1999thermodynamics, dePaula+17}. For all $|\Delta| < 1$, there are spinon excitations with the following dispersion:
\begin{align}
    E_s(k) &= \frac{\pi}{2} \frac{\sin\gamma}{\gamma} \sin{k}.
\end{align}
As before, these are fractionalised excitations and cannot be excited in isolation by a local quench on a chain with an even number of sites \footnote{For odd-length chains this ought to be possible, see e.g. \cite{groha2017spinon} for the construction of the corresponding excitation spectrum for the Heisenberg model.} (see Ref.~\cite{Kulka+25} for how to construct them). We would therefore expect spinons to be excited in pairs. Any observable that couples to pairs of them would lead to a broad 2-spinon continuum in the QSF, similar to the previous case of the AFM. For $0 < \Delta < 1$, the Bethe ansatz also contains string excitations of length $n$, which can be interpreted as bound states of the underlying elementary excitations~\cite{dePaula+17}. Their dispersion is given by:
\begin{align}
    E_{n}(k) &= \frac{\pi J \sin\gamma}{\gamma} \left| \sin \frac{k}{2} \right| \sqrt{1 + \cot^2{\frac{n(\pi/\gamma - 1) \pi}{2}} \sin^2{\frac{k}{2}}},
\end{align}
where the allowed values of $n$ are defined by $\cos\big(\pi /({n+1})\big)<\Delta<1$. In the limit of $\Delta \to 1$, the $n=1$ result approaches the magnon dispersion at the isotropic ferromagnetic point, given by $E(k)=J(1 -\cos k)$, leading to this excitation being labelled `magnon-like' or `spin-wave-like' by some authors~\cite{takahashi1999thermodynamics}. 

A natural approach would be to proceed as before and prepare an approximate ground state, then subsequently probe its excitations via a local quench. However, this strategy becomes challenging in the gapless regime. In contrast to the gapped phases, where the ground state displays area law entanglement and admits an efficient low-depth representation, the XY phase exhibits entanglement that grows logarithmically with system size~\cite{Vidal2003entanglement,latorre2004groundstate}.
As a result, accurately representing the ground state involves increasing the bond dimension of the initial MPS, which translates into greatly increased circuit depth. While nonetheless possible, approximately preparing an initial state would use a substantial fraction of the total circuit depth that is feasible on near-term quantum computers, severely limiting our ability to study the post-quench dynamics beyond all but the shortest evolution times.

\begin{figure}[!tb]
    \centering
    \includegraphics[width=\linewidth]{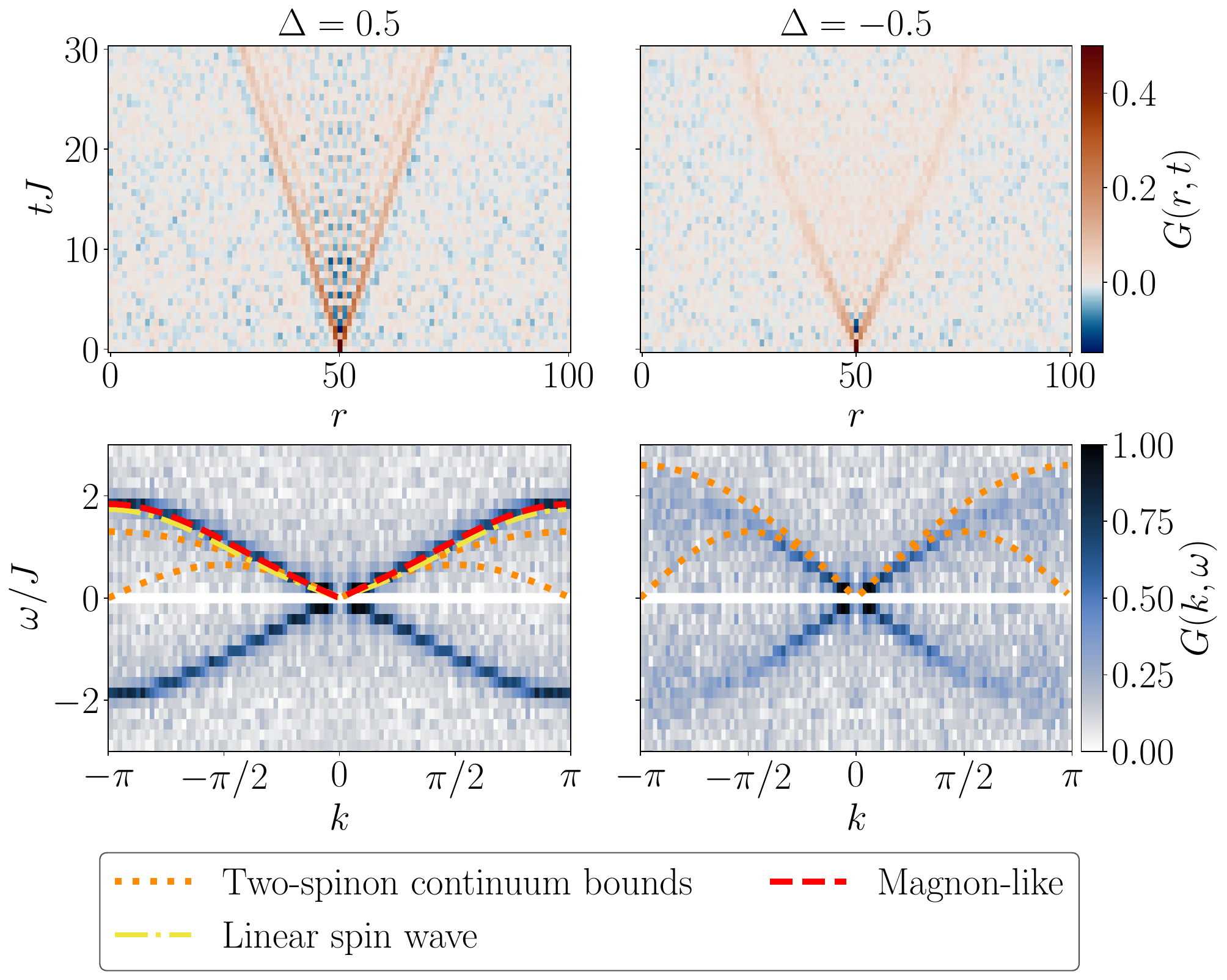}
    \caption{Combined local plus global quench spectroscopy in the XY regime with a $R_y(\pi/2)$ single-site quench, showing real-space dynamics $G(r,t)=2\langle S_z(r,t) \rangle$ (top) and corresponding quench spectral functions $G(k,\omega)$ (bottom) for $\Delta=0.5$ and $\Delta=-0.5$. The orange lines indicate the upper and lower boundaries of the two-spinon continuum (Eq.~\ref{eq:two_spinon_energy}).}
    \label{fig:xy}
\end{figure}

These considerations motivate an alternative approach that avoids explicit ground-state preparation. The principle of quench spectroscopy does not require exact preparation of a given state, it only requires that certain matrix elements between the target manifolds be non-zero (Eqs~\ref{eq:coupling_condition1} and \ref{eq:coupling_condition2}). We can imagine starting in a product state that has overlap with the ground state manifold, and perform a local quench as before. This essentially corresponds to an underlying \emph{global quench} -- as the initial homogeneous state is now no longer even approximately an eigenstate -- immediately followed by a local quench. Global quantum quenches from a variety of initial (matrix) product states have been studied numerically in e.g. Refs~\cite{PhysRevLett.102.130603,barmettler2010quantum,fagotti2014relaxation,collura2015quantum,collura2020order}, particularly in relation to local relaxation in integrable models~\cite{fagotti2013stationary,wouters2014quenching,pozsgay2014correlations,mestyan2015quenching,ilievski2015complete}. Global quenches are characterized by depositing a finite energy density relative to the ground state into the system by means of imposing a particular initial state. In global quenches a light-cone effect occurs in two-point functions in the presence of translational symmetry, while one-point functions exhibit a decay from the initial to their stationary values. While this will strongly modify the matrix elements in Eq.~{\ref{eq:coupling_condition2}}, this does not change the selection rules in Eq.~{\ref{eq:qsf}}.

There has been very little work on local quenches on top of global quenches in the literature, see however Ref.~\cite{fagotti2022global}. This is a highly complex scenario, with both a `double quench' and a model that hosts multiple types of gapless excitation, and so the question is whether quench spectroscopy can tell us anything about the nature of the excitations underlying the observed dynamics. We now show that indeed it can, and that quench spectroscopy of suitably chosen observables on the background of a global quench can reveal the nature and properties of long-lived excitations over the initial state. In particular this can provide access to the spectrum of our target excitation \emph{without} requiring the preparation of a high-fidelity approximation of the ground state, thereby opening a new regime for quench spectroscopy methods~\cite{SM}.

It is useful to distinguish between two settings. First, for finite systems the amount of energy deposited by the global quench can be moderate, i.e. $\Delta E\sim J$.
The initial state after the global quench can then be well approximated by a coherent superposition of the ground state and low-lying excitations. The local quench then generates additional excitations on top, but as low-energy states have the same local structure as the ground state the quench spectral function will be very close to the one obtained by a local quench on top of the ground state. This mechanism allows for the determination of the dispersion relations of elementary excitations in the ground state sector. Second, in situations where the dynamics supports long-lived quasiparticle excitations over the initial state quench spectroscopy will give access to their properties even in the thermodynamic limit.

We implement the global plus local quench protocol by initializing the system in a fully $x$-polarized product state, $\ket{\rightarrow}$, and applying a local perturbation $R_y(\tfrac{\pi}{2})$ at the centre of the chain. 
Our chosen global quench remains in the total spin $S_z=0$ symmetry sector, and by itself generates excitations that do not couple to local observables, as the quench does not break translation invariance. The subsequent local quench involves a superposition of both the $S_z=0$ and $S_z=1/2$ sector, which breaks translation invariance, allowing local probes to couple to excitations. Loosely speaking, one can classify the post-quench dynamics into two distinct behaviours: i) symmetry restoration in the $XY$-plane, where dynamics act to rapidly restore the $U(1)$ symmetry broken by the initial state, and ii) the dynamics of the single spin pointing out of the plane, which is protected by $S_z$ conservation and acts as a long-lived stable excitation.

The results are shown in Fig.~\ref{fig:xy}, for two different values of $\Delta = \pm 0.5$. For $\Delta=0.5$ the quench spectral function exhibits a feature that resembles a single-particle excitation, while for $\Delta=-0.5$ there is additional spectral weight arising from multi-particle excitations. We observe that the dominant feature for both cases is well-described by the magnon-like excitation over the ground state. As noted above this is a consequence of the energy deposited by the global quench being moderate ($2.475J$ $\Delta=0.5$ and $12.625J$ for $\Delta=-0.5$). For $\Delta=0.5$ we also show results obtained by linear spin-wave theory, which becomes quantitatively accurate in the vicinity of $\Delta=1$ and short and intermediate times, see e.g. Ref.~\cite{senese2024out} and references therein. At late times spin wave theory becomes invalid due to the melting of the magnetic order \cite{collura2020order} and the associated finite lifetime of spin wave excitations. While this effect is already significant at $\Delta=0.5$ (see Supplementary Information~\cite{SM}), the spin wave result is seen to still be in good agreement with the observed signal.
The agreement between our two theoretical descriptions is due to the fact that the magnon-like excitation over the ground state reduces to the spin wave for $\Delta\approx 1$. 
In the absence of the exact Bethe ansatz solution, the spin wave analysis could be improved upon by the excitation ansatz~\cite{ExcitationAnsatz1,ExcitationAnsatz2,ExcitationAnsatz3} to compute excitations above a lower energy (higher bond dimension) initial approximation to the ground state, however it is not required in the present case.
While the spin wave result gives a reasonable fit to the observed signal, the magnon-like excitation is consistently a better fit to the data for $0 < \Delta < 1$~\cite{SM}. The data suggests that any linear spin wave excitations in this regime are considerably weaker than signal resulting from the magnon-like excitation.

While it is also possible for the QSF to display algebraic divergences at frequencies $\omega/J=2v \sin(k/2)$ (where $v$ is the excitation velocity)~\cite{villa2020local} that are unrelated to the elementary excitation spectrum, this is not a match for the observed signal shown here. We have also verified using classical MPS simulations (see Ref.~\cite{SM}) the following two points. i) If a weak anisotropy is added in the XY plane, a gap opens in the QSF. This is consistent with how the elementary excitations in this phase should behave, further ruling out the possibility that the observed signal stems from the gapless algebraic divergence of the QSF. ii) When initialising the system in a better approximation to the ground state and computing various observables, the signals in $\langle S^{y}(r,t) \rangle$ and $\langle S^{z}(r,t) \rangle$ are virtually identical to when the $x$-polarised initial state is used, with only the signal in $\langle S^{x}(r,t) \rangle$ differing significantly. Taken together, we find that despite not starting in the true ground state of the system, we are nonetheless able to extract the excitation spectrum corresponding to the magnon-like string excitation for the system sizes and global quenches considered. This is an exciting new development for quench spectroscopy, as it indicates that initial states prepared within subspaces of the correct symmetry sector (with quantum numbers $\sum_r\langle S^{y}(r,t) \rangle = 0$ and $\sum_r \langle S^{z}(r,t) \rangle=0$, in this case) are still able to host stable excitations that can be measured spectroscopically by observables that couple to these subspaces. By contrast, the observable that couples to the direction in which the correct symmetry was broken (with quantum number $\sum_r\langle S^{x}(r,t) \rangle \neq 0$) is unable to unveil the correct excitation spectrum. Elsewhere in the gapless regime, for $\Delta<0$, we expect that linear spin wave theory is not valid and only the two-spinon excitations exist, and indeed we recover a signal consistent with the lower boundary of the two-spinon continuum for $|k| \lesssim \pi/2$.

\section*{Discussion}

Our results have established quench spectroscopy on digital quantum computers as a fast, flexible, and versatile probe of the collective properties of many-body quantum matter, going beyond previous work on the topic to even include situations where elementary excitation spectra can be measured without the need to first prepare a complex ground state. The procedure of quench spectroscopy is similar to earlier proposals to directly measure the dynamical structure factor on quantum computers~\cite{Knap+13,Baez+20,Lee+26}, however with a few notable differences. The main difference is that, rather than reconstructing the dynamical structure factor via linear response theory, we directly compute a different quantity---the quench spectral function---which encodes the same spectral features but requires no linear-response assumption, widening its regime of validity to include initial states far from equilibrium. The quench spectral function provides similar information to established experimental procedures such as inelastic neutron scattering~\cite{Gao+24}. In contrast to other recent work using digital quantum computers for spectroscopy which computed static structure factors~\cite{Leclerc+26} or leveraged classically compiled time-evolved states to increase the accessible evolution times~\cite{Lee+26}, here we focus on dynamical probes where all time evolution is performed directly on the quantum hardware, with our use of approximate quantum compilation (for initial state preparation only) restricted to the case of the AFM. 

The configurability of digital quantum devices makes them an ideal platform for quench spectroscopy, and our results show that this procedure is remarkably robust to the errors and noise present in current generation quantum computers. Performing quench spectroscopy on models with known excitation spectra acts as a highly non-trivial benchmark for the quantum computers themselves, as this is a probe of collective, emergent excitations. By computing spectral functions and comparing with neutron scattering data, we can verify that quantum computers are able to realise target models of interest, and from there further properties of interest such as entanglement entropy may be measured directly on quantum hardware without requiring complex reconstruction from neutron scattering data~\cite{Scheie+25}.
Notably, we have shown that precise preparation of---or knowledge of---the ground state is not required, provided the initial state lies in the correct symmetry sector and deposits only a moderate energy density relative to the ground state, the elementary excitation spectrum can be recovered even from straightforward, easily-prepared product states. 
This is a new avenue for quench spectroscopy, and motivates future work to fully establish the possibilities and limitations of this approach. We have suggested a general principle based on symmetry by which such states may be identified in the future.

With the accuracy of quantum simulators for spectroscopy in one dimension now established, our work opens the doorway towards characterising the excitation spectra of complex quantum matter beyond one dimensional spin chains, making use of the connectivity afforded by quantum simulators to pursue systems with long-range couplings or two-dimensional layouts. Among other things, this will allow us to conduct spectral studies of exotic phenomena that can appear in two dimensions, such as quantum spin liquids~\cite{Savary+17,Nandkishore+21,Lancaster23}, or novel van der Waals magnets~\cite{Burch+18,Park+26}, where entanglement properties can be measured much more directly on quantum simulators than in conventional neutron scattering experiments.
This approach will also extend to regimes where approximations such as linear spin wave theory break down~\cite{Scheie+23,Despres+24}. While challenging to directly realise bosonic systems on current digital quantum computers, it would be interesting to extend this approach used here to study dimerised spin-$1/2$ systems whose excitations can be described as hard-core bosons, leading to unusual forms of magnetic order stemming from Bose-Einstein condensation of excitations~\cite{Giamarchi+08,Zapf+14}, where inelastic neutron scattering is a key probe of their behaviour~\cite{Ruegg+03, Merchant+14}. In parallel with this, it will also be important to continue to develop powerful classical techniques for long-time dynamics in two dimensions~\cite{Thomson+24}, in order to benchmark and verify the quantum hardware and better establish where the regime of practical quantum advantage may lie.

Our results establish quench spectroscopy as a practical spectroscopic tool on present-day quantum hardware. On a single superconducting processor, using only lightweight error mitigation, we have been able to selectively resolve magnons, multi-magnon bound states, and two-spinon continua across the phase diagram of a 101-site XXZ spin chain. We have further shown that relevant spectral features can in some cases be resolved using a reference state away from equilibrium, thus substantially broadening the scope of quench spectroscopic studies on quantum simulators. As quantum processors continue to improve in scale and fidelity, this approach offers a fast and flexible way to measure excitation spectra in regimes that are difficult to access using conventional means, such as two-dimensional strongly correlated systems, models with long range interactions, and novel phenomena beyond the reach of standard equilibrium spectroscopy.

\section*{Methods}
\label{sec:method}
\subsection*{Local quench spectroscopy protocol}

Local quench spectroscopy aims to resolve energy differences between eigenstates of a quantum system by computing the dynamics of a local observable ${O}({\bf r},t)$ following a local quench out of the ground state. In our case, we restrict to cases where the quench is a $\pi/2$ rotation of a single spin or a pair of spins in the middle of the chain.
In a very general form, the space–time signal that we measure is given by:
\begin{align}
G({\bf r},t) &= \langle {O}({\bf r},t) \rangle =  \textrm{Tr}[\rho_0 {O}({\bf r},t)],
\end{align}
where $\rho_0$ is the initial density matrix immediately following the quench, $t$ is the time since the quench, and ${\bf r}$ is the spatial location (which could in principle be in dimensions higher than one). Making use of translation invariance of the Hamiltonian, we can write this in the common eigenbasis of the Hamiltonian $H$ and momentum operator ${\bf P}$ as:
\begin{align}
G({\bf r},t) &=  \sum_{n,n'} \rho_0^{n'n}
e^{i(E_n - E_{n'})t}
e^{i({\bf P}_{n'} - {\bf P}_n) \cdot {\bf r}}
\langle n | {O} | n' \rangle,
\end{align}
showing that the oscillations in space and time encode energy and momentum differences between eigenstates. We now define the object that allows us to precisely resolve these differences.

As mentioned in the main text, the quench spectral function (QSF) is given by:
\begin{align}
G({\bf k},\omega) &= \int dr\, dt \, e^{-i({\bf k} \cdot {\bf r} - \omega t)} G({\bf r},t) \nonumber \\
&= (2 \pi)^2 \sum_{n, n'} \rho_0^{n,n'} \braket{n' | {O}| n} \nonumber \\
& \quad \quad \times \delta({\bf P}_{n} - {\bf P}_{n'} - {\bf k}) \delta(E_n - E_{n'} - \omega) 
\label{eq:qsf2}
\end{align}
which exhibits peaks at energies $\omega = E_{n} - E_{n'}$ and momenta ${\bf k} = {\bf P}_{n} - {\bf P}_{n'}$.
As the spectrum is obtained by Fourier transforming the real-time signal, obtaining a target frequency resolution $\Delta \omega$ requires evolution to corresponding times $T \sim1/\Delta\omega$.

In order for Eq.~\ref{eq:qsf2} to be non-zero, two conditions must be satisfied~\cite{villa2020local}:
\begin{align}
\langle n' | {O} | n \rangle & \neq 0, \label{eq:coupling_condition}\\
\rho_0^{n, n'} = \braket{n | \psi_0} \braket{\psi_0 | n'} &\neq 0.
\end{align}

Up to this point, the formalism allows us to study transitions between any arbitrary pair of states. If we are interested in weak quenches which perturb the ground state $\ket{0}$ and generate low-lying excitations only, we can set $\ket{n'} = \ket{0}$ to be the ground state in order to pick out transitions between the ground state and target excited state manifold, and we can write the initial state as a local operator ${L}$ acting on the vacuum, which we can in turn expand in the basis of eigenstates as $\ket{\psi_0} = {L} \ket{0} = \sum_m c_m \ket{m}$ (where we can assume the sum over states includes only low-lying states. We can then write the vacuum state as $\ket{0} = {L}^{\dagger} \ket{\psi_0} =\sum_m c_m  {L}^{\dagger} \ket{m}$.

With this in mind, the first of the above definitions becomes $\sum_m c^{*}_m \braket{m| {L} {O}|n}$, from which we can see that one practical (but not unique) choice to fix both ${O}$ and ${L}$ is ${O}={L}^\dagger$, i.e. we measure an observable that `reverses' the quench. To satisfy the second condition, we require that the quench operator ${L} \propto \mathbbm{1} - {Q}$ where ${Q}$ couples the ground state to the target manifold such that $\braket{n| {L} | 0} \propto \braket{n| {Q} | 0} \neq 0$ and $\braket{0| {L}^{\dagger} |0} \propto \braket{0|0} \neq 0$. In the case of spin chains, as we consider here, this is satisfied by a spin rotation operator ${L}=R_y(\pi/2)\propto \mathbbm{1}-i{S^y}$, in turn implying that measuring $\braket{S^y}$ across the chain is sufficient to resolve the spectrum. 
For the study of bound states in the ferromagnetic regime, the above considerations generalise straightforwardly as this is still a local quench and a local observable (albeit both are defined on two neighbouring sites, rather than on a single one). 

The overall experimental procedure can therefore be described succinctly by the following steps.

\vspace{0.1in}
\textbf{Protocol:}
\begin{enumerate}
\item Prepare an initial state $|\psi_0\rangle$
\item Apply a local quench ${L}$
\item Evolve under ${H}$ for time $t$
\item Measure a local observable ${O}$
\item Fourier transform $(r,t)\rightarrow(k,\omega)$
\end{enumerate}

The quench prepares a superposition of eigenstates with different energies and momenta. Under time evolution, their relative phases give rise to oscillatory contributions to local observables at frequencies and wavevectors set by eigenstate energy and momentum differences. The Fourier transform resolves these contributions as spectral features in the QSF.

\subsection*{Ground state preparation}

In the AFM phase, our classical MPS simulations~\cite{SM} indicated that use of a N\'eel product state as the initial state was sufficient to resolve the two-spinon continuum, but that the observable signal was rather weak for $\Delta = -2.5$, likely making it difficult to measure on quantum hardware. Additionally, we found that the resulting QSF exhibited additional signals beneath (but very close to) the continuum associated with the algebraic divergence mentioned in the main text. Our MPS simulations indicated that a stronger signal and a cleaner result could be obtained by starting in an initial state that was closer to the true ground state.
In this case, we use DMRG to find a classical MPS approximation to the ground state, followed by the approximate quantum compilation to turn the classical MPS into a shallow circuit which approximately prepares the ground state. More specifically, given a target state $\ket{\psi}$ and a variational ansatz ${U}(\vec{\theta})\ket{0}$, the goal is to minimise the fidelity-based cost function:
\begin{equation}
    C=1-|\bra{\psi}{U}(\vec{\theta})\ket{0}|^2.
    \label{eq:aqc_cost_function}
\end{equation}
We use AQC-Tensor~\cite{robertson2025approximate, qiskit-addon-aqc-tensor}, an open-source Qiskit Addon, which uses tensor networks to evaluate the cost function and its gradients. 

Since the XXZ model is gapped in the AFM phase, we can efficiently obtain MPS representations of its ground states using the DMRG algorithm. We use these MPS as the target states, $\ket{\psi}$ in Eq.~\ref{eq:aqc_cost_function}. Due to the gapped nature of the Hamiltonian, the ground states exhibit short-ranged, exponentially decaying, correlations~\cite{nachtergaele2006lieb}. For this reason, we use a 1D nearest-neighbour $\mathrm{SU}(4)$ brickwork ansatz, due to its ability to support short range correlations. Specifically, $L$ brickwork layers are able to support non-zero correlations $\langle S^z_iS^z_j\rangle-\langle S^z_i\rangle\langle S^z_j\rangle$ between sites separated by up to $|i-j|=4L-1$. Using this relationship, we can estimate that a brickwork circuit of $(\xi+1)/4$ layers should be necessary to capture a state of correlation length $\xi$, up to exponentially small correlations. This is a \textit{lower bound}, since an ansatz which can support correlations over a certain length scale may not be expressive enough to represent \textit{any} state with correlations on that length scale. While it is possible to exactly represent MPS using sequential quantum circuits~\cite{schon2005sequential, schon2007sequential, lin2021real, ran2020encoding,mpstocircuit2025}, the $\mathcal{O}(N)$ scaling of circuit depth is often not necessary for gapped phases with short-range correlated ground states due to the lack of long-range entanglement and correlations.
For the compilation procedure, we used the \texttt{quimb}~\cite{gray2018quimb} backend for the tensor network computations in tandem with the \texttt{adam}~\cite{kingma2017adam} optimizer.

\subsection*{Error suppression and mitigation}

We applied several standard error-suppression and mitigation techniques available within the Qiskit Runtime framework~\cite{qiskit2024,kim2023evidence}.
Firstly, we applied Pauli twirling to both gate operations and measurements~\cite{wallman2016noise,hashim2021randomized}. Pauli twirling randomizes coherent hardware errors by conjugating circuit operations with randomly chosen Pauli operators whose net action cancels in the absence of noise. Averaging over these randomizations converts coherent over- and under-rotations into an effectively stochastic noise channel, which produces more stable expectation values with repeated sampling. The resulting noise has a remarkably flat Fourier spectrum~\cite{Foldager+23,Dalzell+24,Sun+25}, and so the Fourier transform acts like a filter, separating the signal (with a well-defined Fourier spectrum) from the noise (distributed evenly across the Fourier transformed signal). To reduce readout errors, we used twirled readout error extinction (TREX)~\cite{vandenberg2022model}. TREX measures the response of the readout channel under randomized Pauli-frame transformations and uses this information to and correct systematic measurement biases. We performed the TREX calibration procedure using 500 randomizations. For Pauli twirling, we likewise used 500 randomizations, with the runtime determining the number of shots per randomization automatically.
Additionally, we applied dynamical decoupling using the XpXm pulse sequence~\cite{viola1999dynamical}. This suppresses dephasing arising from unwanted coupling to the environment by applying precise, timed sequences of control pulses during inactive portions of the evolution.

Where suitable, we also use physics-model-motivated error mitigation by employing an antisymmetric quench protocol. Rather than measuring the dynamics following only a single local rotation $R_y(\theta)$, we separately performed experiments with both $R_y(+\theta)$ and $R_y(-\theta)$ quenches and constructed the antisymmetrised response:
\begin{equation}
G_{\mathrm{asym}}(r,t) = \frac{1}{2}\left[G_{+\theta}(r,t)-G_{-\theta}(r,t)\right].
\end{equation}
This procedure removes contributions that are even under $\theta \rightarrow -\theta$, including static backgrounds, readout offsets, and symmetric components of the post-quench dynamics unrelated to the direction of the local quench. This procedure isolates the response directly associated with the injected local excitation. In practice, this substantially enhances the visibility of the dispersive spectral features in the resulting quench spectral function. We ensure both experiments are executed within the same Qiskit Runtime \texttt{Session} for to maximize cancellation of hardware drifts and correlated noise.

We did not employ Zero Noise Extrapolation, probabilistic error amplification, post-selection, or extrapolation-based error cancellation in the results shown in the main text. The spectral features arise directly from experimentally measured dynamics using only lightweight mitigation techniques whose sampling overhead remains independent of system size.

\subsection*{Measurement and Fourier analysis}

We measure the local space-time dynamics, $G(r,t)$, for all spatial locations along the chain. This takes the form of expectation values of one-local or two-local observables (depending on the specific experiment), which we measure using the Qiskit \texttt{Estimator} primitive.
We then extract the QSF as the space-time Fourier transform, $G(k,\omega) = \int dr\, dt \, e^{-i(kr - \omega t)} G(r,t)$ using the discrete Fast Fourier transform as implemented in \texttt{NumPy}~\cite{numpy2020}. 

Since the model and ground state are symmetric around the central spin, the dynamics should also be symmetric. For this reason, we average the expectation values around the central spin, as $G(j_c+r,t)\rightarrow\frac{1}{2}(G(j_c+r,t)+G(j_c-r,t))$, prior to performing the Fourier transform. In the case of the two-site quench, where we measure $\langle S^i_xS^{i+1}_x - S^i_yS^{i+1}_y\rangle$, we symmetrise around the central bond.

\acknowledgments

This work was supported by the Hartree National Centre for Digital Innovation, a UK Government-funded collaboration between STFC and IBM, as well as the Engineering and Physical Sciences Research Council (EP/Z533518/1 and UKRI4257). Data and code are available at~\cite{data}.

\section*{Contributions}

D.A.M. and S.J.T. contributed equally to this work. The initial project was conceived by S.J.T. and D.A.M., with later input from A.G.G. and F.H.L.E. G.W.P, D.A.M., N.T.M.S. and A.G.G. contributed to the circuit structure and experimental design. All quantum circuits were run by D.A.M, with input from G.W.P. and S.B. G.W.P. and D.A.M. performed the circuit compilation. S.B. made key contributions to achieving high performance on the available hardware. Classical simulations were performed by S.J.T. and D.A.M. F.H.L.E. advised on Bethe ansatz and linear spin wave theory, and suggested investigation of bound states. Data were analysed and interpreted by D.A.M., S.J.T., G.W.P., F.H.L.E. and A.G.G. S.J.T. and D.A.M. wrote the manuscript, with input from G.W.P., F.H.L.E and A.G.G. All authors discussed the results and the final manuscript.

\bibliography{refs}

\end{document}


\newacronym{aqc}{AQC}{approximate quantum compiling}
\newacronym{mps}{MPS}{matrix product state}
\newacronym{dmrg}{DMRG}{density matrix renormalisation group}
\glsdisablehyper

\title{Supplementary Information to ``Quench Spectroscopy of Magnetic Excitations on a Superconducting Quantum Processor"}

\author{D.~A.~Millar\,\orcidlink{0000-0003-3713-8997}}
\email{declan.millar@ibm.com}
\affiliation{IBM Research, Hursley, Winchester, SO21 2JN, United Kingdom}
\affiliation{School of Physics and Astronomy, University of Southampton, Southampton SO17 1BJ, UK}

\author{G.~W.~Pennington\,\orcidlink{0009-0008-9257-5878}}
\affiliation{The Hartree Centre, STFC, Sci-Tech Daresbury, Warrington WA4 4AD, United Kingdom}

\author{N.~T.~M.~Siow}
\affiliation{London Centre for Nanotechnology, University College London, Gordon St., London, WC1H 0AH, United Kingdom}

\author{S.~Brandhofer}
\affiliation{IBM Research, Ehningen, 71139, Germany}

\author{J.~Crain\,\orcidlink{0000-0001-8672-9158}}
\affiliation{IBM Research, Hursley, Winchester, SO21 2JN, United Kingdom}
\affiliation{Clarendon Laboratory, University of Oxford, Oxford OX1 3PU, UK}

\author{F.~H.~L.~Essler\,\orcidlink{0000-0002-1127-5830}}
\affiliation{Clarendon Laboratory, University of Oxford, Oxford OX1 3PU, UK}

\author{A.~G.~Green\,\orcidlink{0000-0002-3923-5291}}
\affiliation{London Centre for Nanotechnology, University College London, Gordon St., London, WC1H 0AH, United Kingdom}

\author{S.~J.~Thomson\,\orcidlink{0000-0001-9065-9842}}
\email{steven.thomson@ed.ac.uk}
\affiliation{SUPA, School of Physics and Astronomy, University of Edinburgh, Peter Guthrie Tait Road, Edinburgh EH9 3FD, UK}

\date{\today}

\begin{abstract}
This Supplementary Material contains additional details on the combined global plus local quench procedure, circuit compilation, time-evolution circuits, Trotter error analysis, linear spin wave theory analysis, as well as extensive classical simulations supporting the quantum hardware results shown in the main text.
\end{abstract}

\maketitle
\onecolumngrid

\section{Separable Initial States \& Light Cones}

In the case of a combined global and local quench, the underlying dynamics can be quite complicated. It is important to demonstrate that, in the cases we consider, the underlying global quench does not qualitatively change the physics e.g. by introducing qualitatively different excitations or new effects due to the injection of an extensive amount of energy over the ground state. Our results in the main text demonstrate this empirically, clearly showing that the measured dynamics in the XY regime reproduce the expected elementary excitation spectrum for finite system sizes and finite evolution times despite the initial state not being even approximately the ground state. In Sections~\ref{sec:lsw} and ~\ref{sec:mps}, we show that in the regimes we consider, the underlying dynamics due to the global quench (i.e. the dynamics in the XY plane outside of the local quench light cone) reach an approximate stationary state within relatively short times (Figs~\ref{fig.XY50}, \ref{fig.XY50_GLOBAL}, \ref{fig.XY-50}, and \ref{fig.XY-50_GLOBAL}), i.e. before the light cone from the local quench is able to propagate significantly. This indicates that the dynamics of the `background state' outside of the light cone play no role. Here, we provide an additional analytical argument to underpin these observations, making clear the role of system size and evolution time. 

In the case of separable initial states, such as the one considered in the XY regime of our work, we provide an analytical argument suggesting that the results we find are not affected by the dynamics of the underlying global quench outside of the light cone. This is consistent with our procedure working in the limit of large systems, despite the extensive energy injected by the global quench. The initial expectation value can be written:
\begin{align}
    \langle O(r,t) \rangle = \textrm{Tr}[\rho_0 O(r,t)] \label{eq.trace}
\end{align}
A consequence of the Lieb--Robinson bound is that the Heisenberg-picture time evolution of a local operator is quasi-local. This means that for suitable constants $\mu$, $v_\mathrm{LR}$ and $C>0$ there exists an operator $A_{L,R_t}(r,t)$ supported in a region of radius
$R_t=v_{\mathrm{LR}}|t|+R_0$ around $r$, such that
\begin{equation}
    \left\vert\!\vert O(r,t) - A_{L,R_t}(r,t)\right\vert\!\!\vert\le C |\!| O(r)|\!| e^{-\mu (R_t - v_\mathrm{LR}t)}.
    \label{eq:LR}
\end{equation}
As long as the buffer radius $R_0$ can be taken to be sufficiently large for the right-hand side to be negligible $A_{L,R_t}(r,t)$ approximates the original $O(r,t)$ to any desired precision and we can replace $O(r,t)$ by $A_{L, R_t}(r,t)$ in all expectation values.
In the case where the initial state $|\psi_0\rangle$ is a product state and $A_{L,R_t}(r)$ is supported on $R_t$, the separability of the initial state leads to
\begin{equation}
    \langle\psi_0 \vert O(r,t) \vert \psi_0 \rangle \approx \langle \psi_0 \vert A_{L, R_t}(r,t) \vert \psi_0 \rangle_{R_t},
\end{equation}
where $\vert \psi_0\rangle_{R_t}$ denotes the product state restricted to the region $R_{t}$ where the operator $A_{L, R_t}(r,t)$ acts non-trivially.

In addition, if we specify to the case of the XY phase where our chosen observable is $S^z(r,t)$, the expectation value with respect to both the true ground state and the $x$-polarised state will be zero (as all spins lie in the XY plane) unless the region $R_t$ contains the quench site $j_c$. This is a considerable simplification.

While the above observation is general, we can gain some further analytical insight by specifying to the $x$-polarised initial state. This is separable, as it is given by $\ket{\psi_0} = R_y(\pi/2)^{j_c} \prod_i \frac{1}{\sqrt{2}}(\ket{0} + \ket{1}) = \frac{1}{2^{L/2}} (\ket{0} + \ket{1}) \otimes (\ket{0} + \ket{1}) \otimes (\ket{0} + \ket{1}) \otimes ... \otimes R_y(\pi/2)^{j_c}(\ket{0} + \ket{1}) \otimes ... \otimes (\ket{0} + \ket{1})$. We can split this into three subsystems, denoted $A$, $B=R_t$ and $C$ such that $\rho_0 = \rho^A_0 \otimes \rho^B_0 \otimes \rho^C_0$. Subsystems $A$ and $C$ represent regions where the observable acts trivially and subsystem $B$ covers the region $R_t$ which (by the above argument for our specific observable) must contain the quench site, otherwise Eq.~\ref{eq.trace} evaluates to zero. Using the relation $\textrm{Tr}[A \otimes B] = \textrm{Tr}[A] \textrm{Tr}[B]$ and approximating the action of the time-evolved operator outside of the light cone by the identity, the expectation value becomes:
\begin{align}
    \langle O(r,t) \rangle \approx \textrm{Tr}[\rho^A_0 \mathbbm{1}] \textrm{Tr}[\rho^B_0 A_{L, R_t}(r,t) ] \textrm{Tr}[\rho^C_0 \mathbbm{1}]
\end{align}
Away from the quench site, the initial $x$-polarised state is a uniform superposition of all computational basis states. Performing the traces over subsystems $A$ and $C$, we find that $\textrm{Tr}[\rho^A_0 \mathbbm{1}]  = \textrm{Tr}[\rho^C_0 \mathbbm{1}] = 1$ and the expectation value reduces to the trace over subsystem $B$ only. This follows from breaking each trace in region $A$ and $C$ into a product of traces on individual lattice sites and using:
\begin{align}
    \textrm{Tr}[\frac{1}{2}(\ket{0} + \ket{1})(\bra{0} + \bra{1})] =\frac{1}{2}\textrm{Tr}[ \ket{0}\bra{0} + \ket{1}\bra{0} + \ket{0}\bra{1} + \ket{1}\bra{1}] = \frac{1}{2} (1 + 1) = 1
\end{align}
which is valid because by definition regions $A$ and $C$ do not contain the quench site. We are left with:
\begin{align}
    \langle O(r,t) \rangle \approx \textrm{Tr}[\rho^B_0 A_{L, R_t}(r,t) ] .
\label{sm:rdm}
\end{align}
Essentially, the relevant quantity for determining non-zero matrix elements now becomes the reduced density matrix within region $B$, rather than the full density matrix of the entire system. Here we have been able to perform the trace over the other regions explicitly, as our chosen initial state is a product state. A similar picture holds for more general states where we cannot perform the trace analytically. For more complex states within a matrix product state framework, the reduced density matrix can be obtained straightforwardly from a finite-size system by contracting all sites outwith the light cone into a left or right environment tensor. However, for entangled initial states the reduced density matrix $\rho^B$ will be $L$-dependent.

As a consequence of the separability of the initial state the only remaining $L$-dependence in (\ref{sm:rdm}) arises through the approximating operator $A_{L,R_t}(r,t)$. For sufficiently large $L$ this dependence becomes negligible, but the value of $L$ required is not known \emph{a priori}. For our system size $L\approx 100$, as long as $R_t=v_\mathrm{t} + R_0 \ll L$, we expect finite size-corrections to be small, suggesting our results should remain valid in the thermodynamic limit.

Comparing the expected signal from the true ground state $\langle O(r,t) \rangle_{GS}$ with that of our $x$-polarised product state $\langle O(r,t) \rangle_{X}$, we can write the difference in the observed signal as:
\begin{align}
    \langle O(r,t) \rangle_{X} - \langle O(r,t) \rangle_{GS} \approx \textrm{Tr}[(\rho^{B}_X - \rho^{B}_{GS}) A_{L, R_t}(r,t) ] \label{eq.error}
\end{align}
where the approximation is accurate up to the exponentially small factor set by Eq.~\ref{eq:LR}. The difference in signals is therefore bounded by $\vert \rho^B_{X} - \rho^B_{\mathrm{GS}}\vert \cdot \vert A_{L, R_t}(r)$, so it depends on how close the reduced states are and how sensitive the operator is to that difference.
At short times, the first factor is an easier condition to satisfy than requiring overlap of the initial product state and the true ground state over the spatial extent of the entire system, which we would expect to be $\propto 2^{-L}$, compared to the larger $\sim2^{-|B|}$ overlap one would naively expect if only the lattice sites within the light cone contribute. The role of the operator is discussed below.

\section{The Role of Symmetries}

From Eq.~\ref{eq.error}, we can see that the difference between reduced density matrices plays a key role. One could use e.g. the Cauchy-Schwarz inequality to bound this quantity from above by something proportional to the Hilbert-Schmidt distance between these reduced density matrices. This gives rise to a loose upper bound, and crucially misses out the important role that the operator plays. Regardless of how large $(\rho^{B}_X - \rho^{B}_{\mathrm{GS}})$ is, if the operator $O(r,t)$ (or the corresponding approximate operator $A_{L, R_t}(r)$) is not sensitive to this difference (i.e. if it lives in a different symmetry sector), then the trace in Eq.~\ref{eq.error} can still be small or zero. This illustrates that the reduced density matrices need not be similar in every way (which is a highly demanding criterion), only in the way that couples to the target operator chosen to measure the spectrum.

To illustrate this point, let us consider the symmetries present in this particular situation. The Hamiltonian and the $x$-polarized state are invariant under a global $\pi$ rotation around the $x$-axis, $R=R_x(\pi) = \prod_{j=1}^L e^{-i\pi S^x_j}$.

Under this rotation, the three spin components transform in the following ways:
\begin{equation}
R\,S^x_j\,R^\dagger = S^x_j,\qquad
R\,S^y_j\,R^\dagger = -S^y_j,\qquad
R\,S^z_j\,R^\dagger = -S^z_j .
\end{equation}

Conjugation by $R$, written $\mathcal{R}[A]=RAR^\dagger$, is involutory, i.e., $\mathcal{R}^2[A] = R^2 A R^{\dagger 2} = A$, since $R^2 = (-1)^L\mathbbm{1}$ is a scalar multiple of identity and therefore drops out of the conjugation operation. Its eigenvalues are therefore $\pm1$, and any operator splits uniquely into even and odd parts,
\begin{equation}
A = A_+ + A_- ,\qquad A_{\pm} = \tfrac12\left(A \pm RAR^\dagger\right),\qquad RA_\pm R^\dagger = \pm A_\pm .
\end{equation}
Our observables $O(r) = S^y, S^z$ are odd under $R$, and since $[R,H]=0$ so are the time-evolved operators $O(r,t)$. It can also be shown that the trace of any odd operator vanishes,
\begin{equation}
\mathrm{Tr}[A_+\, O(r,t)] = \mathrm{Tr}[R A_+ O(r,t) R^\dagger] = \mathrm{Tr}[(R A_+ R^\dagger)(R O R^\dagger)] = \mathrm{Tr}[A_+(-O(r,t))] = -\mathrm{Tr}[A_+\, O(r,t)] = 0.
\end{equation}
Thus, only the odd part of the state contributes to the signal, $\langle O(r,t)\rangle = \mathrm{Tr}[\rho O(r,t)] = \mathrm{Tr}[\rho_- O(r,t)]$.
The symmetry therefore enters in two ways: it determines what generates a signal in $S^z$ following the local quench, and why the product state reproduces the ground-state result.

Firstly, the local quench $L = R^{j_c}_y(\pi/2)$ breaks the $R$ symmetry, rotating the $x$-moment at the quench site $j_c$ into the (odd) $S^z$ channel, $L\sigma^x_{j_c}L^\dagger \propto \sigma^z_{j_c}$. By definition, the post-quench state expectation value is $\langle S^z_r(t)\rangle_{\mathrm{post}} \equiv \langle L^\dagger S^z_r(t) L\rangle_{\mathrm{pre}}$. Since the pre-quenched state gives $\langle S^z_r(t)\rangle_{\mathrm{pre}} = 0$ for all $r$ by the symmetry $R$ established above, we are free to subtract it. This can be made explicit by writing
\begin{equation}
\langle S^z_r(t)\rangle_{\mathrm{post}} = \langle L^\dagger S^z_r(t) L\rangle_{\mathrm{pre}} = \langle L^\dagger S^z_r(t) L\rangle_{\mathrm{pre}} - \langle S^z_r(t)\rangle_{\mathrm{pre}} = \langle L^\dagger [S^z_r(t), L]\rangle_{\mathrm{pre}},
\end{equation}
where we've used $L^\dagger L = \mathbbm{1}$. The commutator $[S^z_r(t), L]$ measures how much the quench operator $L$ (acting at $j_c$) affects the observable $S^z_r(t)$ (measured at $r$). The Lieb--Robinson bound ensures that this commutator is exponentially small when the spatial separation exceeds the causal light cone, $|r-j_c| > v_{\mathrm{LR}}t$, confining the signal to the region $|r-j_c| \lesssim v_{\mathrm{LR}}t$. The signal is therefore confined to the light cone of the local quench. 

Secondly, the difference between the product-state and ground-state signals depends only on the difference of the odd parts of the two reduced density matrices. Before the local quench, the two states differ only in the \emph{even} sector: the $x$-polarized state carries a finite ordered moment $\langle S^x_j\rangle = \tfrac12$, while the ground state has $\langle S^\alpha_j\rangle_{\mathrm{GS}}=0$ for all $\alpha$ due to the symmetries of $H$.
This can be sketched out by examining a single site, where $\rho^j_X - \rho^j_{\mathrm{GS}} = S^x_j$ exactly, which is even under $R$ and therefore invisible to the odd observable.
The local quench allows $(\rho^j_X$ - $\rho^j_{\mathrm{GS}})$ to acquire an odd component within the quench light cone, which couples to the odd observable, however we still expect the difference to be predominantly in the even sector (i.e. confined to the XY plane).

Our numerical results are consistent with this interpretation. Fig.~\ref{fig.XY99_GROUND} and Fig.~\ref{fig.XY99} show almost identical dispersions in $\langle S^y(r,t)\rangle$ and $\langle S^z(r,t)\rangle$ whether the system is initialised in a close approximation to the ground state or in the $x$-polarized state. Significant differences only appear in $\langle S^x(r,t)\rangle$, i.e., the direction along which the $x$-polarized state breaks the U($1$) rotational symmetry of the ground state.

\section{Circuit Compilation}

\begin{table}[h!]
\centering
\begin{tabular}{c c c}
\hline
$\chi_{\max}$ & $|\langle \psi_{\chi_{\max}} | \psi^{\Delta=2.5}_{\chi=64} \rangle|^2$ & $|\langle \psi_{\chi_{\max}} | \psi^{\Delta=5}_{\chi=44} \rangle|^2$ \\
\hline
1 & 0.0156 & 0.348 \\
2 & 0.8220 & 0.987 \\
3 & 0.9805 & 1.000 \\
4 & 0.9930 & - \\
5 & 0.9989 & - \\
6 & 0.9997 & - \\
\hline
\end{tabular}
\caption{Convergence of the DMRG ground state with increasing maximum bond dimension $\chi_{\max}$ for $\Delta=-2.5,5.0$ and $L=101$. The reference state is obtained with $\chi_{\max}=64$ and $\chi_{\max}=44$, respectively. High fidelities are achieved already for $\chi_{\max}\lesssim 6$ and $\chi_{\max}\lesssim 3$, respectively, indicating that the ground state is well approximated by a low-bond-dimension MPS.}
\label{tab:mps_convergence}
\end{table}

To quantify the accuracy of the ground state, we examine the convergence of the DMRG solution with increasing maximum bond dimension $\chi_{\max}$ in Table~\ref{tab:mps_convergence}. We use a reference state with $\chi=64$, which is large enough that truncation errors are essentially negligible. We observe that high fidelities are achieved already for modest bond dimensions, indicating that the ground state can be accurately captured by a low-bond-dimension MPS. Together with the expected exponential decay of connected correlations in the gapped antiferromagnetic phase~\cite{nachtergaele2006lieb}, this motivates the use of shallow circuit ansatze for approximate quantum compilation.

\section{Hardware Implementation}

\begin{table*}
    \centering
    \begin{tabular}{ccccccc}
        \hline
        Experiment & \shortstack{Initial state\\CNOT depth} & \shortstack{Maximum evolution\\time ($tJ$)} & \shortstack{Number of\\time slices} & \shortstack{Number of\\Trotter steps} & \shortstack{Total CNOT\\depth} & Shots\\
        \hline
        FM magnon, Fig. 2(a) & 0 & 30 & 45 & 15 & 93 & 2048\\
        FM bound state, Fig. 2(b) & 0 & 30 & 61 & 15 & 93 & 8192 \\
        AFM continuum, Fig. 3(a) & 24 & 15 & 25 & 10 & 117 & 10,000\\
        AFM continuum, Fig. 3(b) & 12 & 15 & 25 & 10 & 105 & 10,000\\
        XY global+local, Fig. 4 & 0 & 30 & 45 & 15 & 93 & 2048\\
        \hline
    \end{tabular}
    \caption{Circuit details for the experiments shown in the main text.}
    \label{tab:circuit_details}
\end{table*}

We carried out all quantum hardware experiments using the IBM Quantum Runtime \texttt{Estimator} primitive on the \texttt{ibm\_boston} superconducting quantum processor~\cite{qiskit2024}. To implement the dynamics, we employed a second-order Suzuki--Trotter decomposition~\cite{suzuki1976generalized,lloyd1996universal,trotter1959on} of the time-evolution operator, decomposing the nearest-neighbour exchange terms into sequences of native one- and two-qubit gates.
In order to evolve the initial state to a time $t=M\delta t$, we keep the number of Trotter steps, $M$, fixed and vary the Trotter step size, $\delta t$, rather than keeping the step size fixed and varying the number of steps. In this way the overall circuit structure remained approximately unchanged throughout the spectroscopy protocol, allowing us to compare different evolution times under nearly identical hardware conditions.

Table~\ref{tab:circuit_details} summarizes the parameters used in each experiment, including the maximum evolution time, the number of sampled time slices, the number of Trotter steps, the final two-qubit-gate depth and the number of shots. The deepest circuits were required for the antiferromagnetic spectroscopy experiments, where the Trotterized evolution followed the approximate preparation of an entangled ground state.
To characterize the compiled circuits, we quote the two-qubit-gate depth throughout the manuscript. This quantity provides a more meaningful estimate of accumulated hardware noise than the total logical depth, since single-qubit errors remain substantially smaller than two-qubit errors on current superconducting devices~\cite{mckay2023benchmarking}.

We transpiled all circuits using Qiskit's preset pass managers at optimization level 2 with a fixed transpiler seed~\cite{qiskit2024}. To identify contiguous regions of the processor with low two-qubit-gate error rates and short routing distances, we selected initial qubit layouts informed by error-per-layered-gate (EPLG) values obtained from layer-fidelity benchmarking~\cite{mckay2023benchmarking}.
After transpiling the deepest circuit within a given experiment, we reused the resulting layout for all remaining circuits associated with that dataset. This ensured that all measured time slices sampled essentially the same physical region of the processor and experienced comparable calibration conditions. For hardware execution, we transformed all observables into the final transpiled basis using the layout returned by the transpiler.

\section{Time Evolution Circuits}

We use a second order Suzuki-Trotter decomposition~\cite{trotter1959on} to perform the time evolution. We partition our Hamiltonian, Eq.~1 of the main text, into two commuting groups:
\begin{align}
    H_\mathrm{even}&=\sum_{i~\mathrm{even}}H_{i,i+1} \\
    H_\mathrm{odd}&=\sum_{i~\mathrm{odd}}H_{i,i+1},
\end{align}
where $H_{i,i+1}=-J({S}^x_i{S}^x_{i+1}+{S}^y_i{S}^y_{i+1}+\Delta{S}^z_i{S}^z_{i+1})$ and $H=H_\mathrm{even}+H_\mathrm{odd}$. Using this notation, the time evolution operator becomes:

\begin{align}
    U(t)&=e^{-iHt}\\
    &=\left(e^{-iHt/M}\right)^M\\
    &=\left(e^{-i(H_\mathrm{even}/2+H_\mathrm{odd}+H_\mathrm{even}/2)t/M}\right)^M\\
    &\approx \left(e^{-iH_\mathrm{even}t/2M}e^{-iH_\mathrm{odd}t/M}e^{-iH_\mathrm{even}t/2M}\right)^M,
\end{align}

where in the last step we have employed the Suzuki-Trotter approximation, which becomes exact in the limit of large $M$. From here on, we will define $t/M:=\delta t$ and refer to this as the Trotter step size. Since each $H_{i,i+1}$ term in $H_\mathrm{even}$ or $H_\mathrm{odd}$ acts on a distinct pair of qubits, each exponential in the previous expression exactly decomposes into the product of two-qubit operators: 
\begin{equation}
    e^{-iH_{\mathrm{even}}\delta t/2}=\prod_{i~\mathrm{even}}U_{i,i+1}(\delta t/2),
\end{equation}
where $U_{i,i+1}(\delta t)=e^{-iH_{i,i+1}\delta t}$, and an analogous expression for $e^{-iH_{\mathrm{odd}}\delta t}$. Therefore, each Trotter step consists of a layer of two qubit operators, $U_{i,i+1}$, acting on all even-odd qubit pairs resulting in a half-timestep ($\delta t/2$) of evolution, followed by an analogous full-timestep layer acting on all odd-even pairs, followed by another half-timestep layer acting on all even-odd pairs, as shown in Fig.~\ref{fig:trotter_step}. When multiple Trotter steps are concatenated, the trailing half-timestep layer $e^{-iH_\mathrm{even}\delta t/2}$ from one Trotter step can be combined with the leading half-timestep layer from the subsequent Trotter step, in order to reduce depth and gate count. In this way, the circuit structure of the second-order Trotter decomposition is identical to the first-order decomposition, except that the first layer of even-odd pair interactions only evolves by a half-timestep, and there is an additional half-timestep layer at the end.

\begin{figure*}
\begin{quantikz}[baseline=(current bounding box.center)]
 &\gate[4]{e^{-iH\delta t}}& \\
 && \\
 && \\
 &&
\end{quantikz}
$\approx$
\begin{quantikz}[baseline=(current bounding box.center)]
 &\gate[2]{U_{i,i+1}(\delta t/2)}&&\gate[2]{U_{i,i+1}(\delta t/2)}& \\
 &&\gate[2]{U_{i,i+1}(\delta t)}&& \\
 &\gate[2]{U_{i,i+1}(\delta t/2)}&&\gate[2]{U_{i,i+1}(\delta t/2)}& \\
 &&&&
\end{quantikz}
\caption{Circuit representation of a single second-order Trotter step. When concatenating multiple Trotter steps, the trailing half time-step blocks from one Trotter step can be merged with the leading blocks from the subsequent Trotter step as $U_{i,i+1}(\delta t/2)U_{i,i+1}(\delta t/2)=U_{i,i+1}(\delta t)$, to reduce circuit depth.}
\label{fig:trotter_step}
\end{figure*}
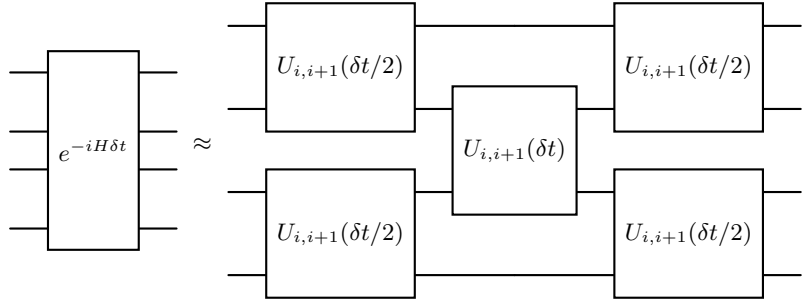

\begin{figure*}
\begin{quantikz}[baseline=(current bounding box.center)]
 &\gate[2]{U_{i,i+1}(\delta t)}& \\
 &&
\end{quantikz}
$=$
\begin{quantikz}[baseline=(current bounding box.center)]
&&\targ{}&\gate{R_z(\theta)}&\ctrl{1}&&\targ{}&\gate{R_z(\pi/2)}& \\
&\gate{R_z(-\pi/2)}&\ctrl{-1}&\gate{R_y(\phi)}&\targ{}&\gate{R_y(\lambda)}&\ctrl{-1}&&
\end{quantikz}
\caption{Circuit representation of the $e^{-iH_{i,i+1}\delta t}$ operator, where $\theta=\pi/2 - J\Delta \delta t/2$, $\phi=J\delta t/2-\pi/2$, and $\lambda=\pi/2-J\delta t/2$~\cite{smith2019simulating, vatan2004optimal}}
\label{fig:cartan_block}
\end{figure*}
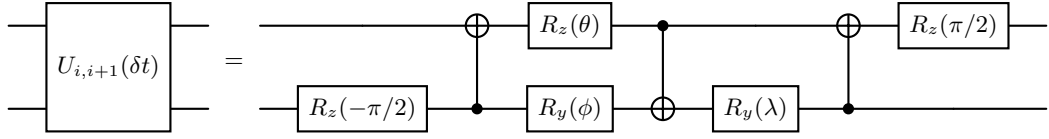

Finally, each $U_{i,i+1}$ term can be decomposed as shown in Fig.~\ref{fig:cartan_block}. It should be noted that we use the same gate conventions as Qiskit and Ref.~\cite{smith2019simulating}, whereas Ref.~\cite{vatan2004optimal} has a relative factor of $-1$ for the angle definition for $R_y$ and $R_z$. Putting all of this together, we have that a second-order Suzuki-Trotter time-evolution circuit for the XXZ Hamiltonian consisting of $M$ Trotter steps has a brickwork structure and a CNOT depth of $6M+3$.

\section{Trotter error for single and bound magnons}

In practice on quantum hardware we exactly simulate evolution under the effective Floquet Hamiltonian $H_F$ given by
\begin{equation}
    U_{\mathrm{Trotter}}(\delta t)^M = e^{-i H_F(\delta t)t}
    \label{eq:floquet}
\end{equation}
where $M$ is the number of Trotter steps, $\delta t$ the step size, and $t=M\delta t$ the total evolution time~\cite{bukov2015universal}.

A single symmetric second-order Suzuki-Trotter step can be written
\begin{equation}
    U_{\mathrm{Trotter}}(\delta t) = e^{-iA\delta t/2}e^{-iB\delta t}e^{-iA\delta t/2}
\end{equation}
the Baker--Campbell--Hausdorff expansion gives
\begin{equation}
    U_{\mathrm{Trotter}}(\delta t) = \exp\left[-iH\delta t - i\delta t^3
    \left(\frac{1}{24}[A,[A,B]] - \frac{1}{12}[B,[B,A]]\right)
    + \mathcal{O}(\delta t^5)\right],
\end{equation}
where $H=A+B$. Comparing exponents for a single step in Eq.~\ref{eq:floquet} gives us
\begin{equation}
    H_F(\delta t) = H + \delta t^2
    \left(\frac{1}{24}[A,[A,B]] - \frac{1}{12}[B,[B,A]]\right)
    + \mathcal{O}(\delta t^4).
\end{equation}
Thus the excitation creates quasiparticles that propagate according to a modified Hamiltonian,
\begin{equation}
    H_F(\delta t) = H + \delta t^2 H^{(2)} + \mathcal{O}(\delta t^4).
\end{equation}
It directly follows that the quasiparticle dispersion will also acquire correction
\begin{equation}
    E_F(k,\delta t) = E_0(k) + \delta t^2 E^{(2)}(k) + \mathcal{O}(\delta t^4),
\end{equation}
and that the group velocity can be written
\begin{equation}
    v_F(k,\delta t) = \frac{\partial E_F}{\partial k} = v_0(k) + \delta t^2 v^{(2)}(k) + \mathcal{O}(\delta t^4).
\end{equation}
Maximizing over momentum, we can write
\begin{equation}
    v^{\max}_F(\delta t) = \max_k \left\vert v_F(k,\delta t)\right\vert = v^{\max}_0 + \delta v\delta t^2 + \mathcal{O}(\delta t^4).
    \label{eq:vFmax}
\end{equation}
The theoretical ballistic light cone edge is given by $r=v^{\max}_0 t$, where $v^{\max}_0$ is the maximum group velocity of the exact model we are attempting to simulate. 
Thus, substituting $\delta t = t/M$ into Eq.~\ref{eq:vFmax}, the effective light-cone edge in our quantum simulation is determined by
\begin{equation}
    r_{\mathrm{edge}} = v^{\max}_0 t + \frac{\delta v}{M^2} t^3 + \mathcal{O}(t^5).
\end{equation}
Hence, the deviation from the ballistic trajectory of the target Hamiltonian is cubic in time and the sign of $\delta v$ determines the direction that the light cone edge bends. If $\delta v < 0$ the cone bends inwards, while $\delta v > 0$ causes the cone to bend outwards.

\subsection{Single magnon}

We first consider the single-magnon excitation,
\begin{equation}
\ket{j}=S_j^+\ket{\Downarrow},
\end{equation}
where $\ket{\Downarrow}$ denotes the fully-polarised ferromagnetic background. In this sector the $S^zS^z$ interaction contributes only a constant energy shift and therefore does not affect the group velocity. Thus we focus on the hopping contribution.

The one-particle Floquet spectrum of the integrable XXZ brickwork circuit is known exactly~\cite{aleiner2021bethe}. In our notation it may be written as
\begin{equation}
\cos\Theta(2k)=\cos^2\alpha-\sin^2\alpha\cos 2k,
\qquad
\alpha=\frac{J\delta t}{2},
\end{equation}
where the factor of $2k$ reflects the doubled unit cell of the even--odd brickwork decomposition. The eigenvalues of the single-step Floquet operator are $e^{-i\Theta(2k)}$, where $\Theta(2k)$ is the phase accumulated during one Trotter step. The corresponding Floquet quasienergy is therefore
\begin{equation}
E_F(k,\delta t)=\frac{\Theta(2k)}{\delta t} = -\frac{2}{\delta t}\arcsin(\sin\alpha\cos k).
\label{eq:magnon_floquet_dispersion}
\end{equation}
where we've chosen the quasienergy branch that reduces to the exact one-magnon dispersion in the $\delta t\rightarrow0$ limit. The Floquet quasienergy may be expanded around small $\delta t$ as
\begin{equation}
E_F(k,\delta t) = -J\cos k + \frac{J^3\delta t^2}{24}\cos k\sin^2 k + \mathcal O(\delta t^4).
\end{equation}
The first term is the usual one-magnon dispersion, while the second term is
the leading Floquet correction. Differentiating,
\begin{equation}
v_F(k,\delta t) = \frac{\partial E_F}{\partial k} = J\sin k + \frac{J^3\delta t^2}{24} \sin k \left(3\cos^2 k-1\right) + \mathcal O(\delta t^4).
\end{equation}
At the maximum-velocity point $k=\pi/2$, this becomes
\begin{equation}
v^{\max}_F(\delta t) = J\left(1-\frac{J^2\delta t^2}{24}\right) + \mathcal O(\delta t^4).
\end{equation}
Comparing with Eq.~\ref{eq:vFmax}, it follows that
\begin{equation}
\delta v_{\mathrm{magnon}} = -\frac{J^3}{24} < 0.
\end{equation}
The Floquet correction therefore suppresses the maximum group velocity of the magnon and causes the magnon light cone to bend inwards.

\subsection{Two-magnon bound state}

The exact two-magnon bound-state dispersion of the target XXZ Hamiltonian in the gapped regime is~\cite{takahashi1999thermodynamics}
\begin{equation}
E(k) = J\, \frac{\sinh\eta}{\sinh 2\eta} \left( \cosh 2\eta-\cos k \right),
\qquad \Delta=\cosh\eta, \qquad \eta>0.
\end{equation}
Using the hyperbolic identity
\begin{equation}
\sinh 2\eta = 2\sinh\eta\cosh\eta = 2\Delta\sinh\eta,
\label{eq:sin2eta}
\end{equation}
this may be written as
\begin{equation}
E(k) = \frac{J}{2\Delta} \left( \cosh 2\eta-\cos k \right).
\end{equation}

The corresponding group velocity is therefore
\begin{equation}
v(k) = \frac{\partial E(k)}{\partial k} = \frac{J}{2\Delta}\sin k,
\end{equation}
which is maximized at $k=\pi/2$. Hence the exact maximum bound-state velocity is
\begin{equation}
v^{\max}_{0} = \frac{J}{2\Delta}.
\end{equation}
The Floquet result should therefore reduce to this value in the $\delta t \to 0$ limit.

The two-magnon bound-state Floquet dispersion of the integrable XXZ brickwork circuit in the gapped regime is also known~\cite{aleiner2021bethe}:
\begin{equation}
\Theta(2k) = \Theta(\pi) + \widetilde{\Theta}(2k),
\end{equation}
with
\begin{equation}
\cos\widetilde{\Theta}(2k) = \frac{\cos^2\alpha\,\sinh^2(2\eta) - \sin^2\alpha\,\sinh^2\eta\,\cos 2k}
{\cos^2\alpha\,\sinh^2(2\eta) + \sin^2\alpha\,\sinh^2\eta}, \qquad \alpha=\frac{J\delta t}{2}.
\end{equation}
Here $k$ is the physical centre-of-mass momentum, while the argument $2k$ reflects the doubled unit cell of the brickwork circuit. The shift $\Theta_b(\pi)$ is momentum independent, so it changes the overall quasienergy offset but does not affect the group velocity. The corresponding Floquet quasienergy is
\begin{equation}
E_F(k,\delta t) = \frac{\Theta(2k)}{\delta t}.
\end{equation}

To simplify the momentum dependence, define
\begin{equation}
A = \cos^2\alpha\,\sinh^2(2\eta), \qquad B = \sin^2\alpha\,\sinh^2\eta.
\end{equation}
The bound-state spectrum may then be written as
\begin{equation}
\cos\widetilde{\Theta}(2k) = \frac{A-B\cos 2k}{A+B}.
\end{equation}
Dividing numerator and denominator by $A$ and defining
\begin{equation}
q = \frac{B}{A} = \frac{\sin^2\alpha\,\sinh^2\eta}{\cos^2\alpha\,\sinh^2(2\eta)},
\end{equation}
gives
\begin{equation}
\cos\widetilde{\Theta}(2k) = \frac{1-q\cos 2k}{1+q}.
\end{equation}

Using Eq.~\ref{eq:sin2eta} we obtain
\begin{equation}
q = \frac{\tan^2\alpha}{4\Delta^2}.
\end{equation}
Since
\begin{equation}
\alpha=\frac{J\delta t}{2},
\end{equation}
it follows that
\begin{equation}
q = \frac{J^2\delta t^2}{16\Delta^2} + \mathcal{O}(\delta t^4).
\end{equation}

Selecting the branch that reduces to the target dispersion as $\delta t\to0$ gives the momentum-dependent bound-state quasienergy
\begin{equation}
E_F(k,\delta t) = \frac{\Theta(\pi)}{\delta t}
- \frac{2}{\delta t}\arcsin\!\left(\sqrt{\frac{q}{1+q}}\,\cos k\right).
\end{equation}
The momentum-independent offset $\Theta(\pi)/\delta t$ plays no role in the group velocity. The momentum dependence is the direct analogue of the single-magnon dispersion Eq.~\ref{eq:magnon_floquet_dispersion}, with $\sin\alpha$ replaced by $\sqrt{q/(1+q)}$.

To obtain the small-$\delta t$ expansion, we use
\begin{equation}
q = \frac{\tan^2\alpha}{4\Delta^2}, \qquad \alpha=\frac{J\delta t}{2}.
\end{equation}
Expanding the square-root factor gives
\begin{equation}
\sqrt{\frac{q}{1+q}} = \frac{\tan\alpha}{\sqrt{4\Delta^2+\tan^2\alpha}}
= \frac{\alpha}{2\Delta} \left[ 1+ \frac{8\Delta^2-3}{24\Delta^2}\alpha^2 + \mathcal{O}(\alpha^4) \right].
\end{equation}
The argument of the arcsine is therefore $\mathcal{O}(\alpha)$, so we must also retain the cubic term in
\begin{equation}
\arcsin x = x+\frac{x^3}{6}+\mathcal{O}(x^5).
\end{equation}
This gives
\begin{equation}
E_F(k,\delta t) = \frac{\Theta(\pi)}{\delta t} - \frac{J}{2\Delta}\cos k
- \frac{J^3\delta t^2}{192\Delta^3} \left[ (8\Delta^2-3)\cos k + \cos^3 k \right]
+ \mathcal{O}(\delta t^4).
\end{equation}
The leading momentum-dependent term reproduces the exact target-Hamiltonian dispersion up to the same constant shift discussed above.

Differentiating gives
\begin{equation}
v_F(k,\delta t) = \frac{\partial E_F}{\partial k} = \frac{J}{2\Delta}\sin k
+ \frac{J^3\delta t^2}{192\Delta^3} \left[ (8\Delta^2-3)+3\cos^2 k \right]\sin k
+ \mathcal{O}(\delta t^4).
\end{equation}
At the maximum-velocity point $k=\pi/2$, this becomes
\begin{equation}
v^{\max}_{F}(\delta t) = \frac{J}{2\Delta} + \frac{J^3\delta t^2}{192\Delta^3} \left(8\Delta^2-3\right)
+ \mathcal{O}(\delta t^4).
\end{equation}
Therefore
\begin{equation}
\delta v_{\mathrm{bound}} = \frac{J^3}{192\Delta^3} \left( 8\Delta^2-3 \right).
\end{equation}

For $\Delta>1$,
\begin{equation}
8\Delta^2 - 3 > 0,
\end{equation}
so
\begin{equation}
\delta v_{\mathrm{bound}} > 0.
\end{equation}
Thus the Floquet correction increases the maximum group velocity of the two-magnon bound state, causing the bound-state light cone to bend outwards.

\section{Linear Spin Wave Theory in the XY Regime}
\label{sec:lsw}

We consider a quench protocol where the initial state takes the form
\begin{align}
|\Psi_0\rangle=\frac{1}{{\cal N}}(\mathbbm{1}+\gamma S^z_{L/2})\ket{\rightarrow} ,
\end{align}
where $\ket{\rightarrow}$ is the ferromagnetic state along the
x-direction, $\gamma$ a free parameter, and ${\cal N}$ is a
normalization factor. This corresponds to a local quench on top of a
global quench. The global part of the quench setup is a particular
case of the ``tilted N\'eel states'' considered in
Ref.~\cite{fagotti2014relaxation}. 

We note that both $H$ and $\ket{\rightarrow}$ are invariant under a
rotation by $\pi$ around the x-direction, which is therefore a
symmetry of the global quench. As $S^z_j$ is odd under this symmetry
we necessarily have $\bra{\rightarrow}S^z_j(t)\ket{\rightarrow}=0$.
The local quench breaks this symmetry and therefore
\begin{align}
{\cal S}(j,t)=\langle \Psi_0|S^z_j(t)|\Psi_0\rangle\neq 0\ 
\end{align}
will exhibit a light cone effect. The question is then what, if
anything, can be inferred from the structure of this light cone.

The energy density is an important parameter. In our case we
have
\begin{align}
\lim_{L\to\infty}\frac{1}{L}\bra{\rightarrow}H\ket{\rightarrow}=-\frac{J}{4}\ ,
\end{align}
while the ground state energy density in the thermodynamic limit is
\begin{align}
e_0(\Delta)=-\frac{J\Delta}{4}-\frac{J\sin\gamma}{2\gamma}\int_{-\infty}^\infty\frac{d\omega}{\omega}\frac{\sinh\big(\frac{(\pi-\gamma)\omega}{\gamma}\big)}{\cosh\omega\sinh\big(\frac{\pi\omega}{\gamma}\big)}\ ,
\end{align}
where $\gamma={\rm arccos(-\Delta)}$. Particular values are
\begin{align}
e_0\left(\frac{1}{2}\right)=-0.2745J\ ,\quad
e_0(0)=-\frac{J}{\pi}\ ,\nonumber
e_0\left(-\frac{1}{2}\right)&=-\frac{3J}{8}\ .
\end{align}
For $L=101$ this implies an energy difference between the ground state
and the initial state of $\Delta E \approx 2.475J$ for $\Delta=\frac{1}{2}$
and $\Delta E=12.625J$ for $\Delta=-\frac{1}{2}$. This ``excess energy'' gets
distributed among excitations over the ground state, and given that
the spectrum is gapless we expect a substantial number of excitations even
though the values of $\Delta E$ are still modest for $L=101$.
\subsection{Short-time behaviour: spin-wave analysis}
\label{sec:spinwaves}
The initial state is fully polarized, and it will take some time $t^*$
before the ordered moment in x-direction decays to zero under time
evolution. The value of $t^*$ depends on the anisotropy $\Delta$ and
becomes large when $\Delta\rightarrow 1$.
This means that the short-time behaviour for $t\ll t^*$ can be
described by spin-wave theory, see e.g. Ref.~\cite{senese2024out}. A
natural conjecture is then that the light-cone observed in ${\cal
  S}(j,t)$ can be described by the spin-wave excitations underlying
the global quench dynamics as long as $1-\Delta$ is small. To work out
the spin-wave dispersion we rotate the spin quantization axis by $\pi
$ around the y-direction, so that the initial state is polarized along
the new z-direction. The Hamiltonian becomes 
\begin{align}
H=-J\sum_{j=1}^L\tilde{S}^z_j\tilde{S}^z_{j+1}+\tilde{S}^y_j\tilde{S}^y_{j+1}+\Delta
\tilde{S}^x_j\tilde{S}^x_{j+1}\ . 
\end{align}
We then use a Holstein-Primakoff representation
\begin{align}
\tilde{S}^z_j=S-a^\dagger_ja_j\ ,\quad
\tilde{S}^+_j=\tilde{S}^x_j+i\tilde{S}^y_j=\sqrt{2S}a_j +\dots
\end{align}
In the linear spinwave approximation we have
\begin{align}
H_{\rm LSW}&=-JS^2L+JS\sum_ja^\dagger_ja_j+a^\dagger_{j+1}a_{j+1}\nonumber\\
  &+JS\sum_j\left[\frac{1-\Delta}{2}a_ja_{j+1}-\frac{1+\Delta}{2}a^\dagger_ja_{j+1}+{\rm h.c.}\right].
\end{align}
Corrections can be straightforwardly accounted for in a
self-consistent mean-field approximation. We now go to momentum space
\begin{align}
a(k)=\frac{1}{\sqrt{L}}\sum_j e^{-ikj}a_j\ ,
\end{align}
and then carry out a Bogoliubov transformation
\begin{align}
a(k)=\cosh\theta_m(k) \alpha_k-\sinh\theta_k\alpha^\dagger_{-k}\ ,\quad \theta_k=\theta_{-k}
\end{align}
to diagonalize the Hamiltonian. 
Setting
\begin{align}
\tanh(2\theta_k)=\frac{\frac{1-\Delta}{2}\cos
  k}{1-\frac{1+\Delta}{2}\cos k}\ ,
\end{align}
we obtain
\begin{align}
H_{\rm
  LSW}=\sum_{k>0}\epsilon(k)\left[\alpha^\dagger_k\alpha_k+\alpha^\dagger_{-k}\alpha_{-k}\right]+{\rm const}
\end{align}
The resulting spinwave dispersion is
\begin{align}
\epsilon(k)=2JS\sqrt{\left(1-\frac{1+\Delta}{2}\cos
  k\ \right)^2-\left(\frac{1-\Delta}{2}\cos k\right)^2}\ .
\end{align}

\subsection{Timescale of Validity of the Linear Spin Wave Approximation}
\begin{figure}
    \centering
    \includegraphics[width=0.7\linewidth]{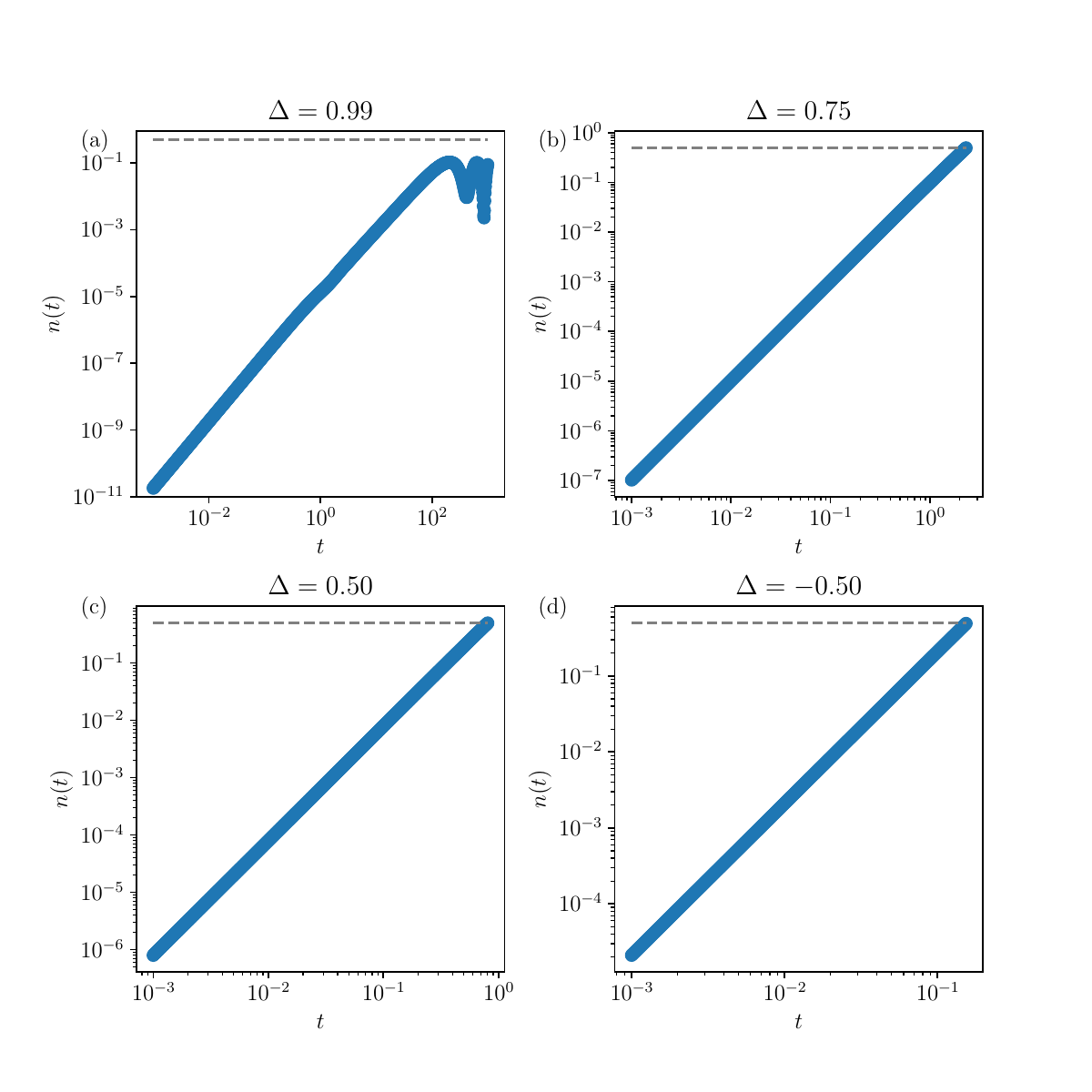}
    \caption{The results of Eq.~\ref{eq.lsw} for four different values of $\Delta$. The grey dashed line in each panel shows the point at which $\langle
0|a^\dagger_j(t)a_j(t)|0\rangle=0.5$, at which point we stop the evolution as this has grown sufficiently large than linear spin wave theory is no longer valid. Note that the timescale required to reach this point depends strongly on $\Delta$, with values close to $\Delta=1$ remaining much more stable to long times, as we would expect. In the case of $\Delta=0.99$ we do not reach this threshold within the time window considered.}
    \label{fig.LSW}
\end{figure}

We have
\begin{align}
\langle
0|a^\dagger_j(t)a_j(t)|0\rangle=\frac{1}{L}\sum_k\sinh^2(2\theta_k)\sin^2\big(\epsilon(k)t\big)\ .
\label{eq.lsw}
\end{align}
As long as this remains small the use of the spin wave approximation is
justified. Tuning $\Delta$ closer to the ferromagnetic point extends
the time window where this remains true. This quantity is shown in Fig.~\ref{fig.LSW} for four values of $\Delta$, confirming that it behaves as expected and giving a quantitative idea of the timescales on which linear spin wave theory is valid. As the timescales are small in most cases, one can Taylor expand the $\sin^2(\epsilon(k) t)$ term in Eq.~\ref{eq.lsw} and at short times all curves grow approximately like $\propto F(k) t^2$, where $F(k)$ is the sum over all $k$-dependent terms in the prefactor of $t^2$ following the Taylor expansion.

\section{Classical MPS Simulations}
\label{sec:mps}

\begin{figure}
    \centering
    \includegraphics[width=\linewidth]{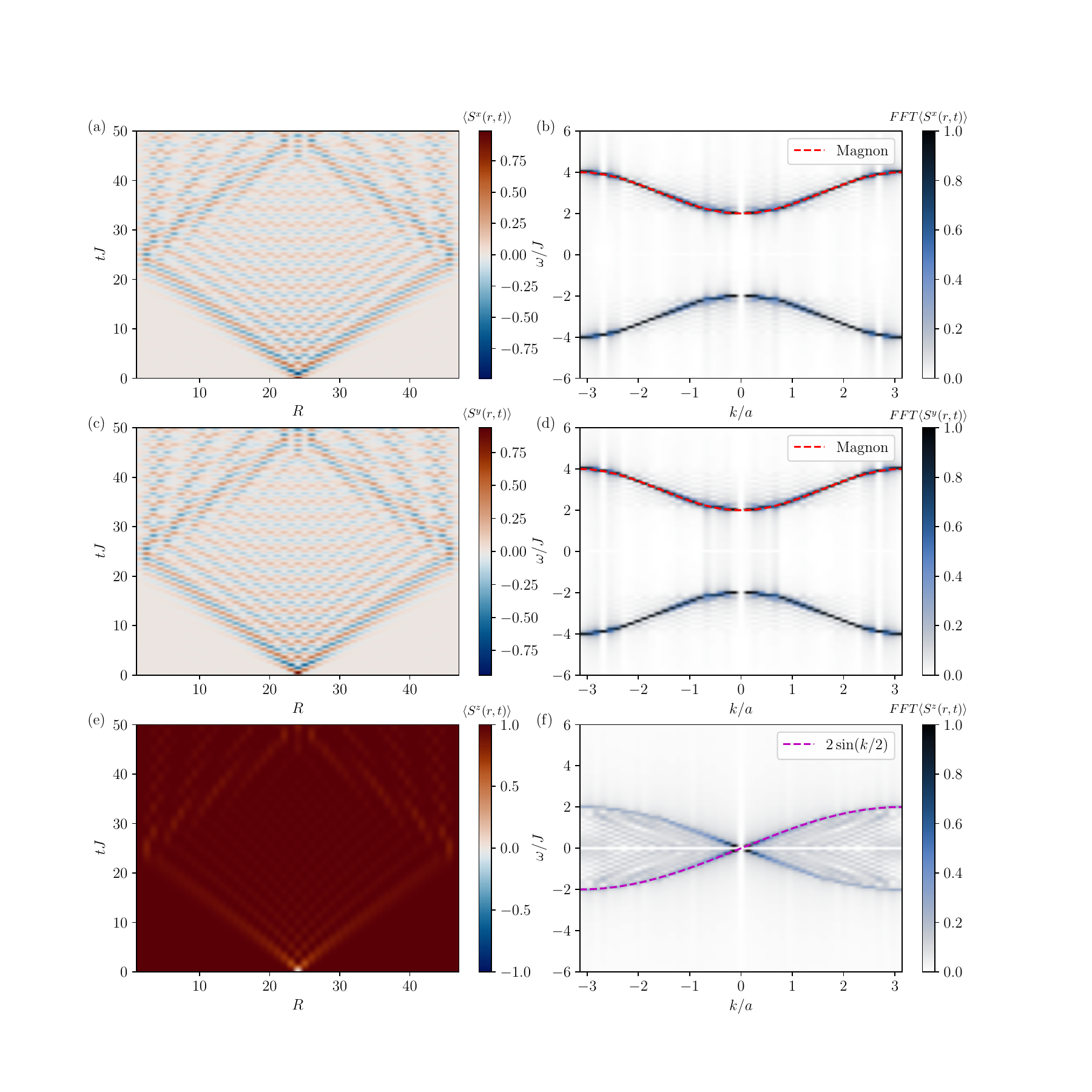}
    \caption{Real-time dynamics following a single-site $R_y(\pi/2)$ quench on the centre lattice site of a length $L=47$ spin chain with $\Delta=3.0$, starting in a fully polarised product sate with all spins in the $z$-direction. Panels (a), (c) and (e) show the dynamics of $\langle S^{x}(r,t) \rangle$, $\langle S^{y}(r,t) \rangle$ and $\langle S^{z}(r,t) \rangle$ respectively, while panels (b), (d) and (f) show the corresponding quench spectral functions. Depending on the observable we measure, we either find a sharp magnon dispersion, or a gapless signal associated with the algebraic divergence of the quench spectral function. One can loosely interpret the physical origin of this signal as being due to the single flipped spin having the same energy cost on any lattice site, therefore it is free to move without paying any additional energy cost.}
    \label{fig.FM3}
\end{figure}

\begin{figure}
    \centering
    \includegraphics[width=\linewidth]{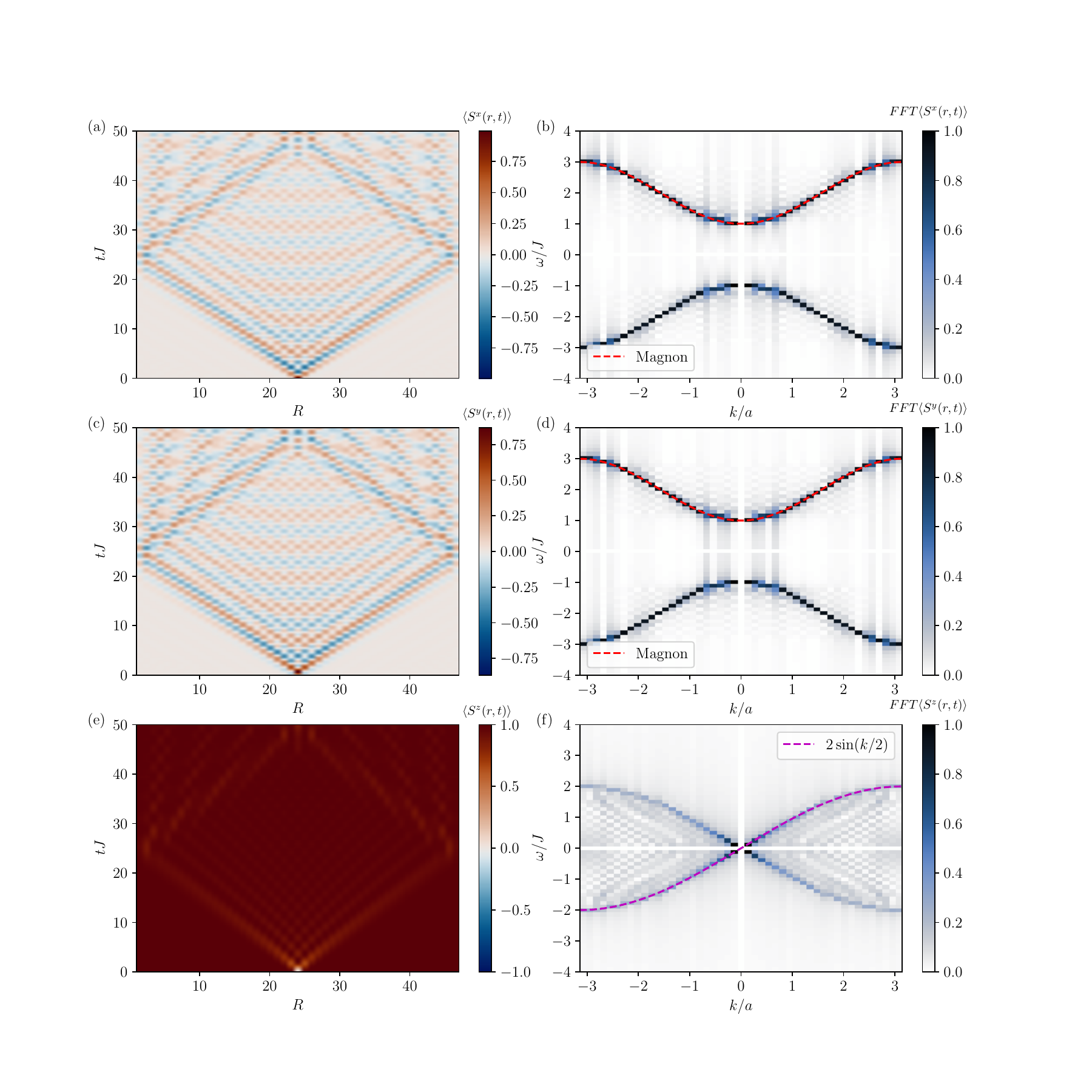}
    \caption{The same as Fig.~\ref{fig.FM3}, but for $\Delta=2.0$. Note that the gap has become smaller, visible in the top two panels by the magnon bands moving closer to zero frequency.}
    \label{fig.FM2}
\end{figure}

\begin{figure}
    \centering
    \includegraphics[width=\linewidth]{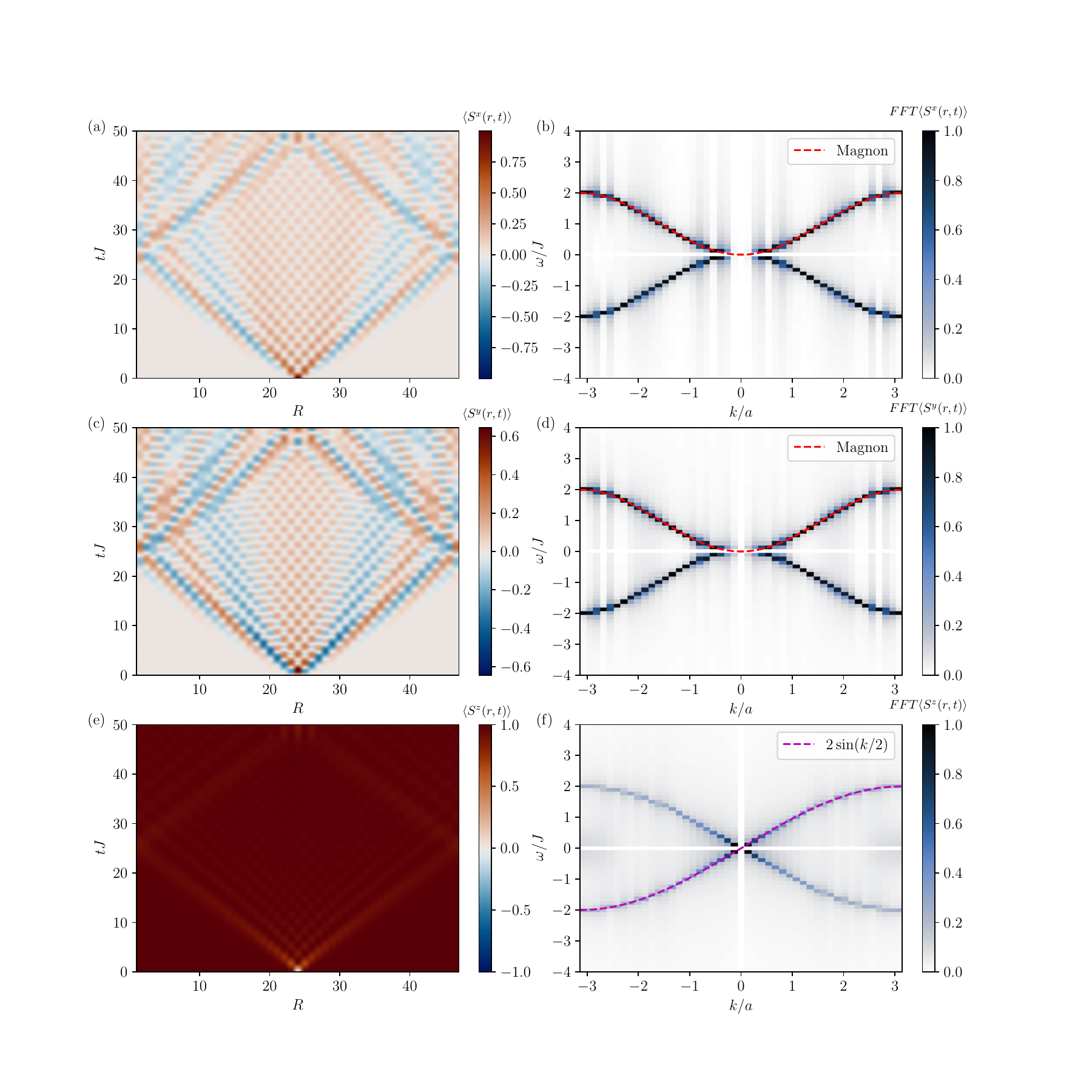}
    \caption{The same as Fig.~\ref{fig.FM3}, but for $\Delta=1.0$. This is the isotopic Heisenberg point, where the magnon excitation becomes gapless.}
    \label{fig.FM1}
\end{figure}

\begin{figure}
    \centering
    \includegraphics[width=\linewidth]{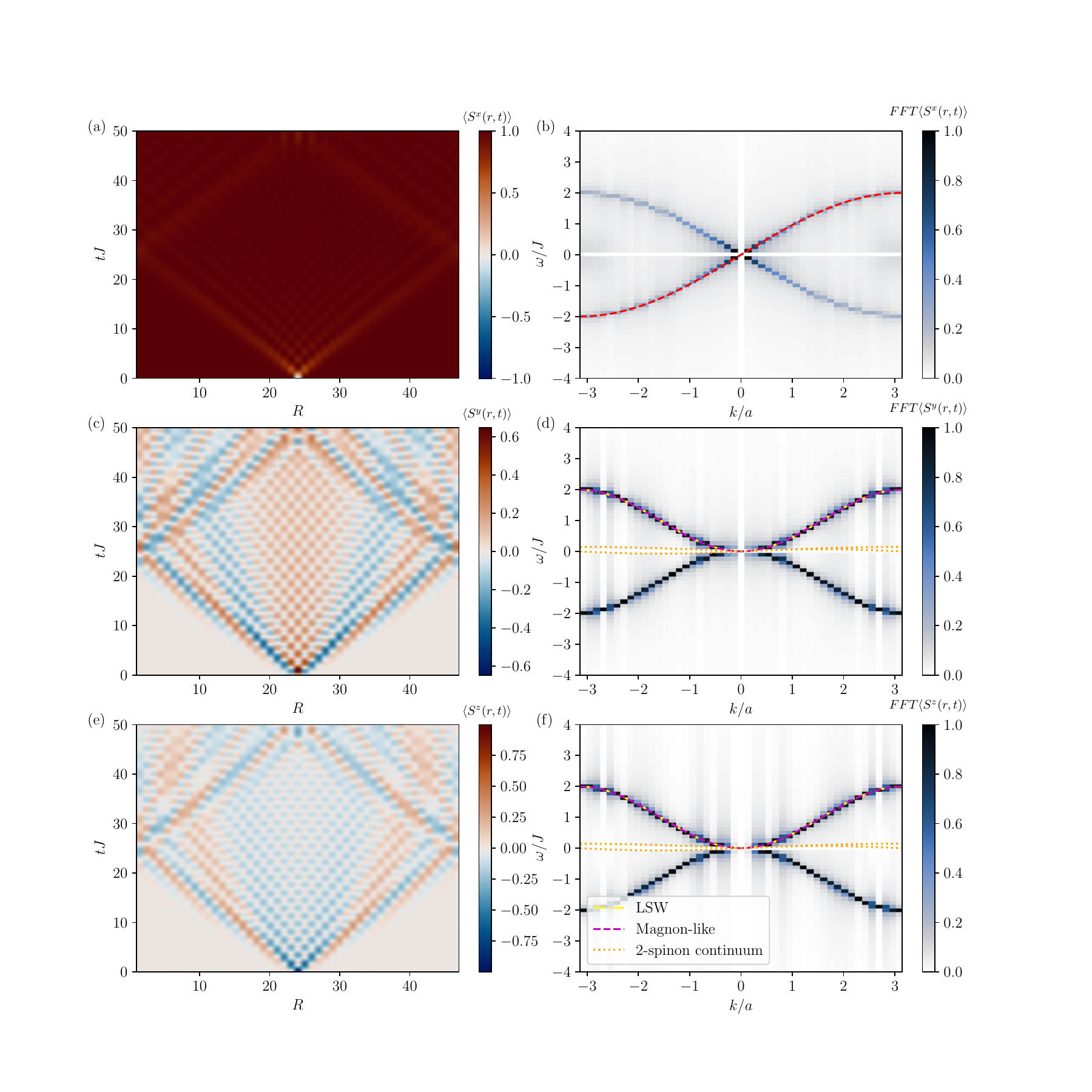}
    \caption{The same as Fig.~\ref{fig.FM3}, but for $\Delta=0.99$ and now we start in an $x$-polarised product state. This point is just within the gapless XY phase, but is sufficiently close to the isotropic $\Delta=1.0$ point that the $x$-polarised state is a very good approximation to an eigenstate. Note that now the gapless signal associated with the algebraic divergence appears in the Fourier transform of $\langle S^{x}(r,t) \rangle$, as this is the direction along which the initial state was ordered. As we are now in the gapless phase, we draw several possible candidate signals coming from the 2-spinon continuum, the magnon-like excitation from the Bethe ansatz, and the solution from linear spin wave theory (LSW). The LSW solution matches the magnon-like solution perfectly. There is no visible signal from the 2-spinon continuum in this regime.}
    \label{fig.XY99}
\end{figure}

\begin{figure}
    \centering
    \includegraphics[width=\linewidth]{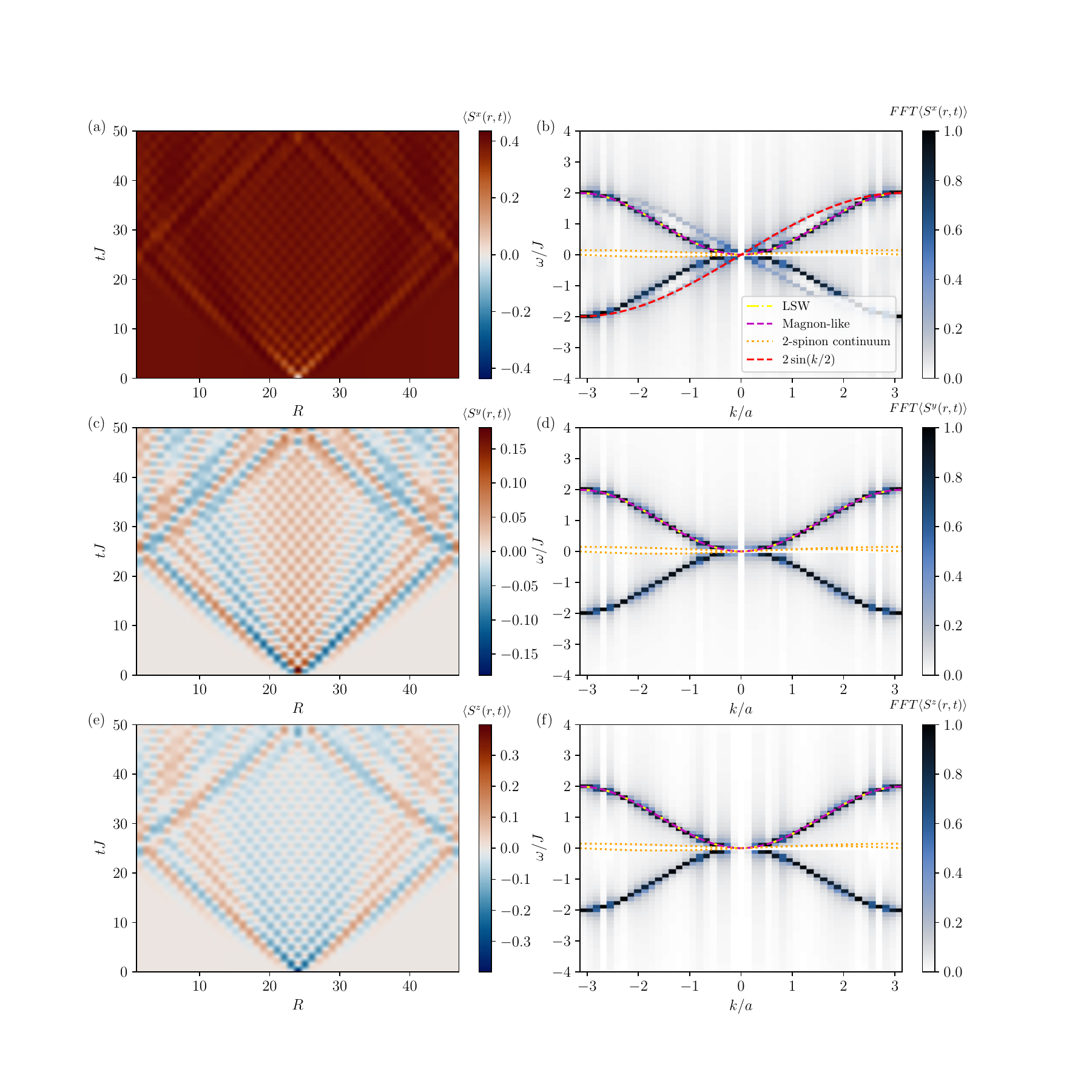}
    \caption{The same as Fig.~\ref{fig.XY99}, but now starting in a closer approximation to the ground state, obtained using DMRG which converged to an initial state with bond dimension $\chi=22$. (Note that the initial state still retains a non-negligible polarisation along the $x$ direction, which is not surprising given that this bond dimension is far too small to obtain the expected $U(1)$ symmetric ground state.) The $\langle S^{y}(r,t) \rangle$ and $\langle S^{z}(r,t) \rangle$ behaviour is largely unchanged, but the $\langle S^{x}(r,t) \rangle$ signal is strongly modified and now contains both the algebraic divergence \emph{and} the magnon-like excitation. This provides a clear insight into why the $x$-polarised state was able to unveil the elementary excitation spectrum despite not being the ground state: the behaviour of the quench in two of the three variables was identical, and only in the direction where the underlying symmetry of the Hamiltonian was explicitly broken did we see a spurious signal unrelated to the elementary excitations. It is interesting to note that the $2 \sin(k/2)$ signal seen in the Fourier transform of $\langle S^{x}(r,t) \rangle$ seems fainter than the signal due to the magnon-like excitation; it is tempting to speculate that this signal would fade entirely for a fully $U(1)$-symmetric ground state.}
    \label{fig.XY99_GROUND}
\end{figure}

\begin{figure}
    \centering
    \includegraphics[width=\linewidth]{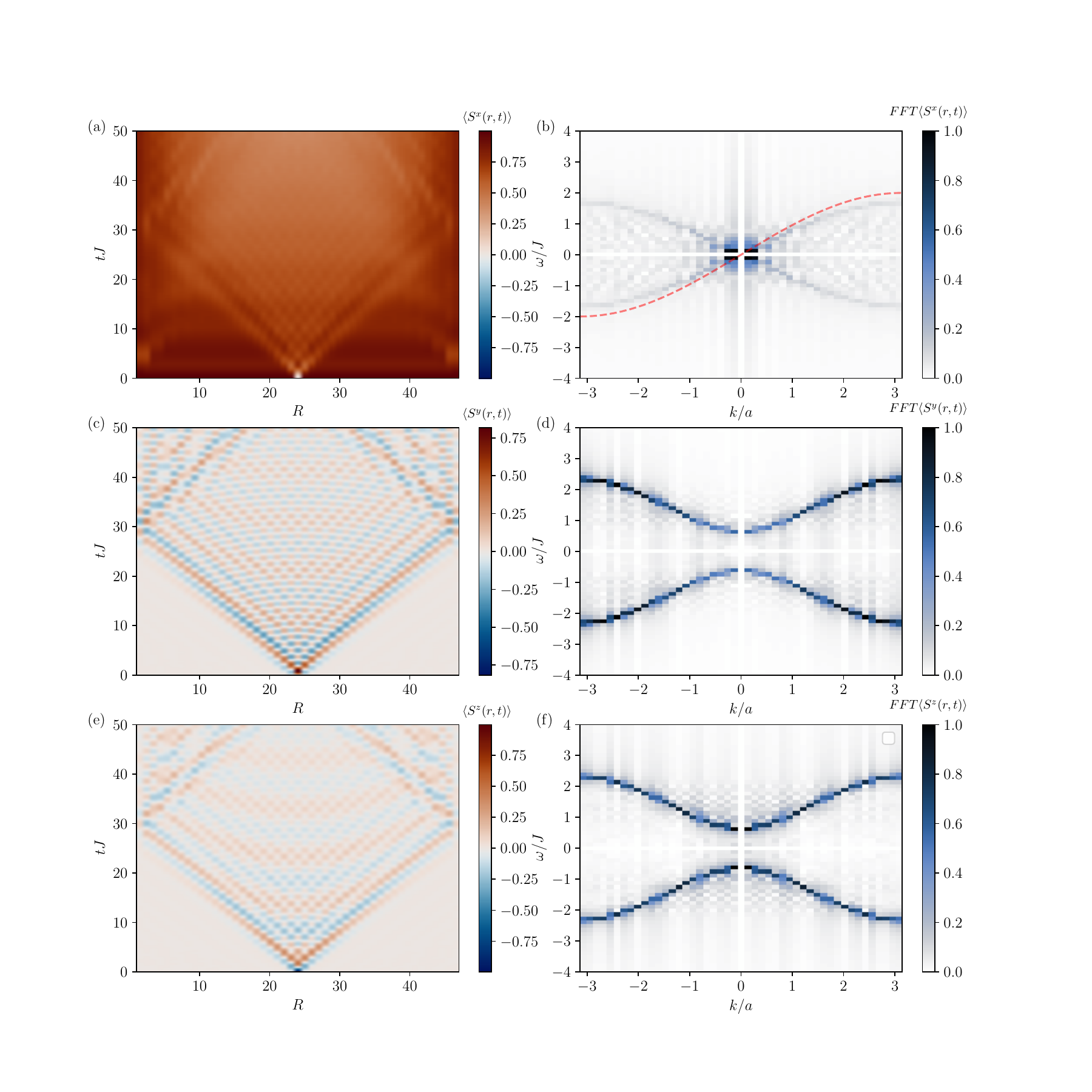}
    \caption{The same as Fig.~\ref{fig.XY99}, but with an added anisotropy in the XY plane, such that the Hamiltonian reads $H = -J\sum_{i=1}^{N-1} \left( (1 + \gamma)S^x_i S^x_{i+1}
+ (1 - \gamma) S^y_i S^y_{i+1}
+ \Delta S^z_i S^z_{i+1} \right)$, and here we use $\gamma=0.5$. The gapless signal associated with the algebraic divergence seen in the Fourier transform of $\langle S^{x}(r,t) \rangle$ is still present, but has been modified slightly from the previous $2\sin(k/2)$ form. In the other observables, a gap has been opened, as we would expect from the elementary excitation spectrum, confirming that this signal indeed originates from the elementary excitations.}
    \label{fig.XY99_ANISOTROPIC}
\end{figure}

\begin{figure}
    \centering
    \includegraphics[width=\linewidth]{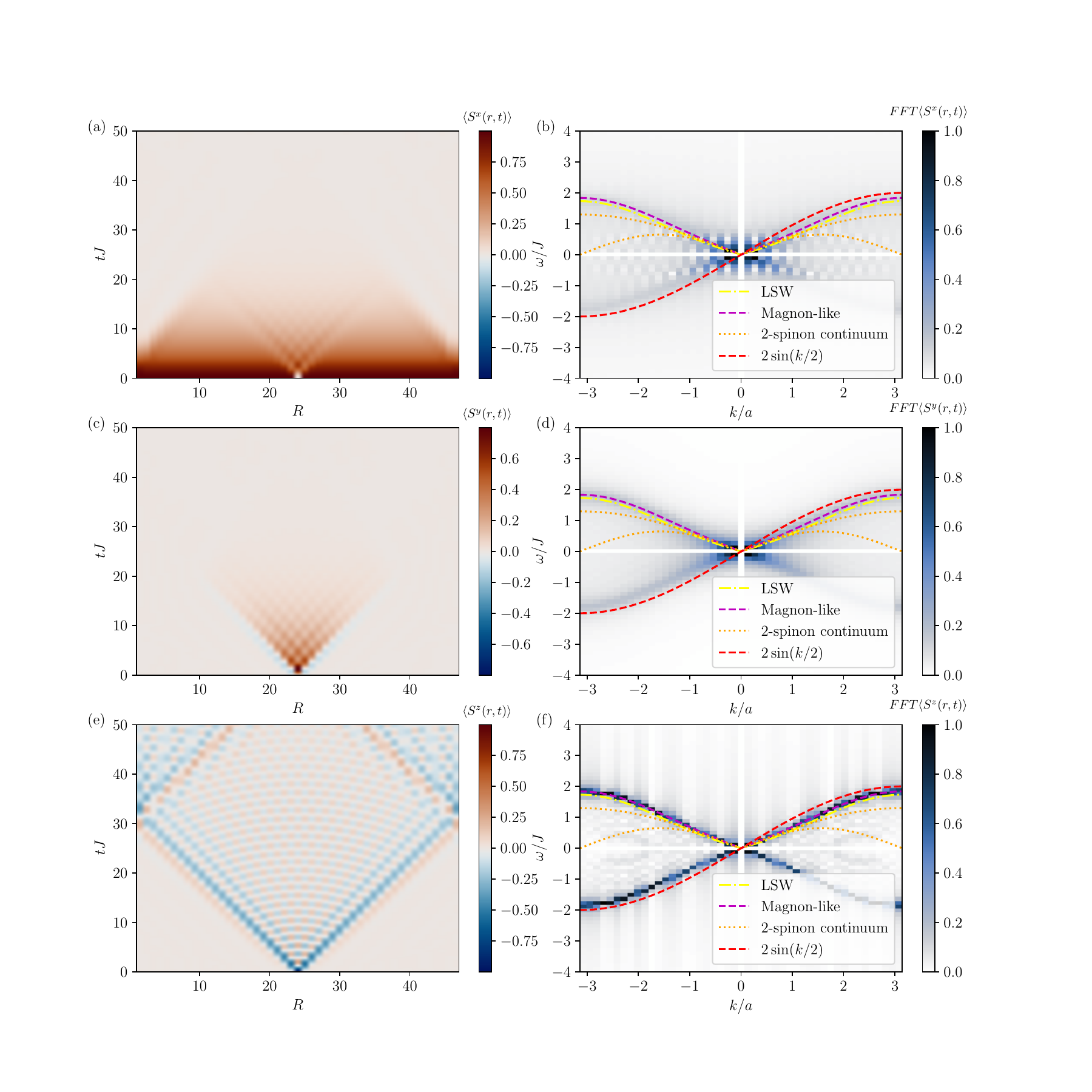}
    \caption{The same as Fig.~\ref{fig.FM3}, but for $\Delta=0.5$, again starting from an $x$-polarised product state. This point is further into the XY phase, and now we can see from $\langle S^{x}(r,t) \rangle$ that the initial ordered moment decays as the dynamics act to restore the $U(1)$ symmetry in the XY-plane. The $\langle S^{z}(r,t) \rangle$ signal, by contrast, is much more robust due to the conservation of $\sum_R \langle S^{x}(r,t) \rangle$ meaning that the excitation is visible to much longer times, giving a sharper Fourier spectrum. In contrast to the $\Delta=0.99$ result, notice that all three observables now display signals consistent with the magnon-like excitation.}
    \label{fig.XY50}
\end{figure}

\begin{figure}
    \centering
    \includegraphics[width=\linewidth]{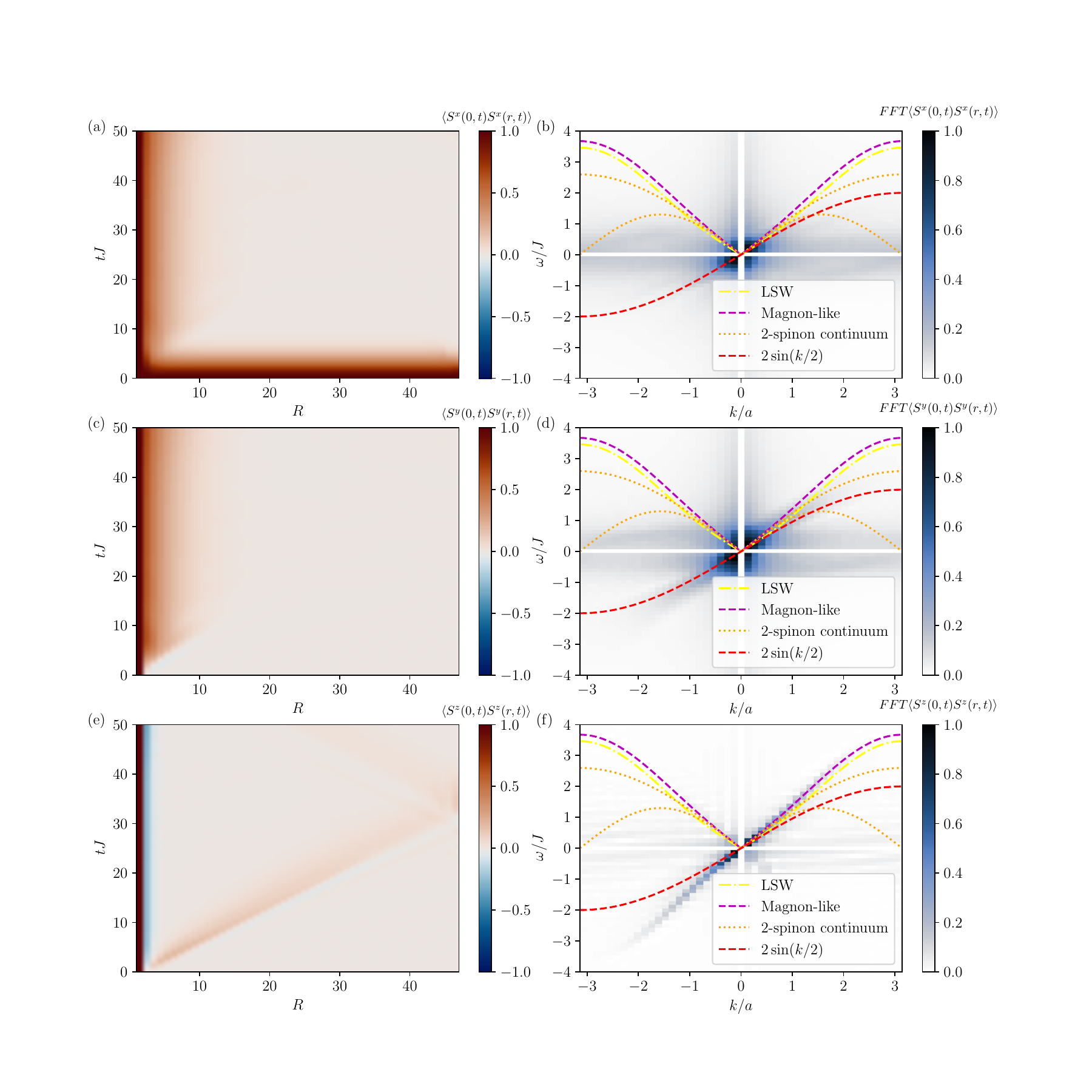}
    \caption{Global quench spectroscopy for $\Delta=0.5$. Here, the initial state is just an $x$-polarised state \emph{without} being followed by a local quench. As a result, the quench does not break translation invariance, and the spectrum cannot be probed by local observables. Instead, we must use two-point (equal time) correlation functions. In addition, we expect to recover a signal that is twice the spectrum, rather than the spectrum itself. For details of the method, see Ref.~\cite{villa2019unraveling}. The Fourier transforms of $\langle S^{x}(r,t) \rangle$ and $\langle S^{y}(r,t) \rangle$ do not show a clear spectrum, but the un-transformed variables give an insight into the physics. We can clearly see the `melting' of the initial order in the $x$-direction, and the formation of correlations in the $y$-direction. By contrast, in the $z$-direction we again find a signal that perfectly matches the expected signal for the magnon-like excitation.}
    \label{fig.XY50_GLOBAL}
\end{figure}

\begin{figure}
    \centering
    \includegraphics[width=\linewidth]{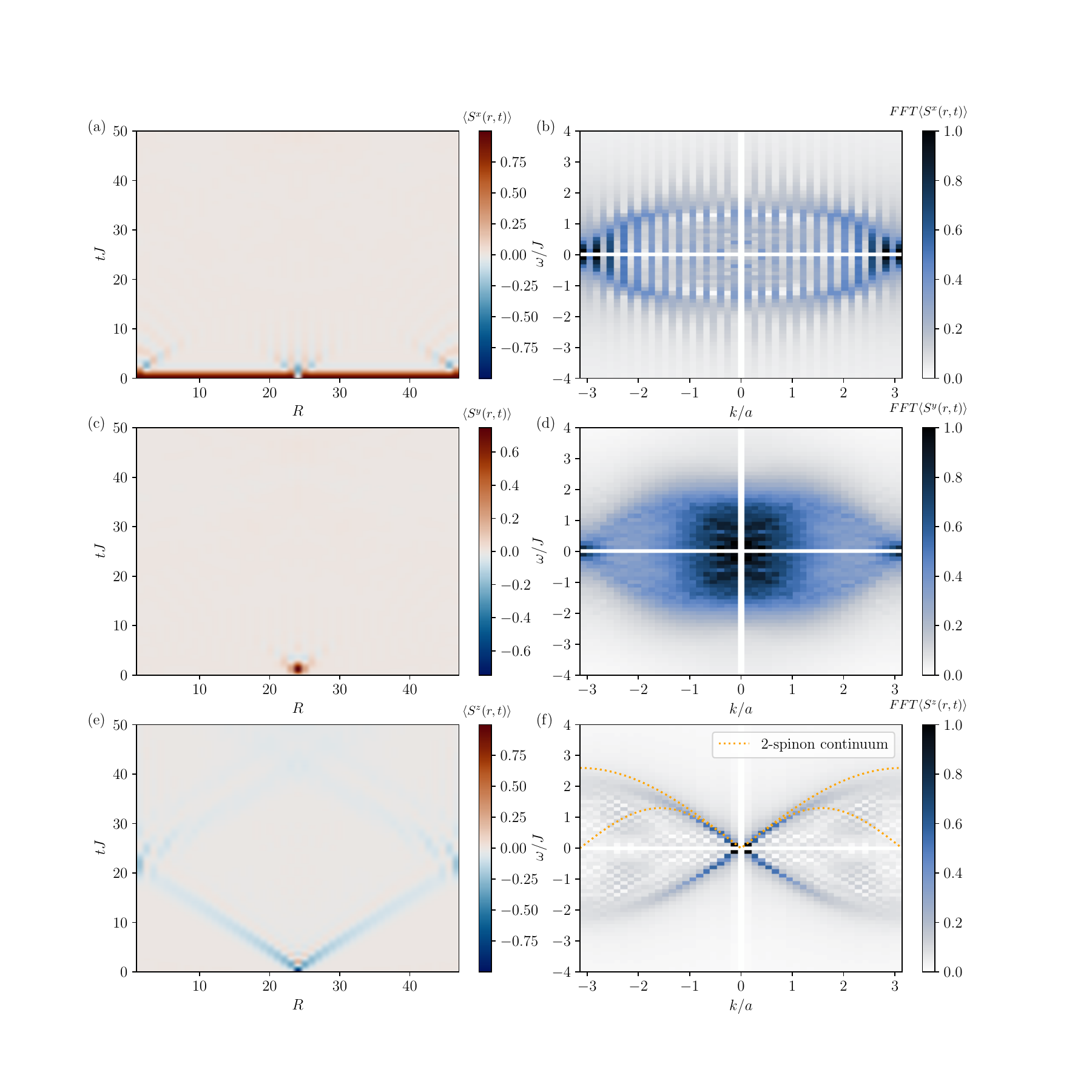}
    \caption{The same as Fig.~\ref{fig.XY50}, but for $\Delta=-0.5$, again starting from an $x$-polarised product state. While the Fourier features for $\langle S^{x}(r,t) \rangle$ and $\langle S^{y}(r,t) \rangle$ appear quite dramatic, note that the extremely rapid melting of the initial order means that the real-space signal is non-existent after a very short time, and so the Fourier transform is dominated by small fluctuations rather than elementary excitations. The Fourier transform of $\langle S^{z}(r,t) \rangle$ shows behaviour consistent with the expected continuum and very similar to the result obtained from the quantum hardware. }
    \label{fig.XY-50}
\end{figure}

\begin{figure}
    \centering
    \includegraphics[width=\linewidth]{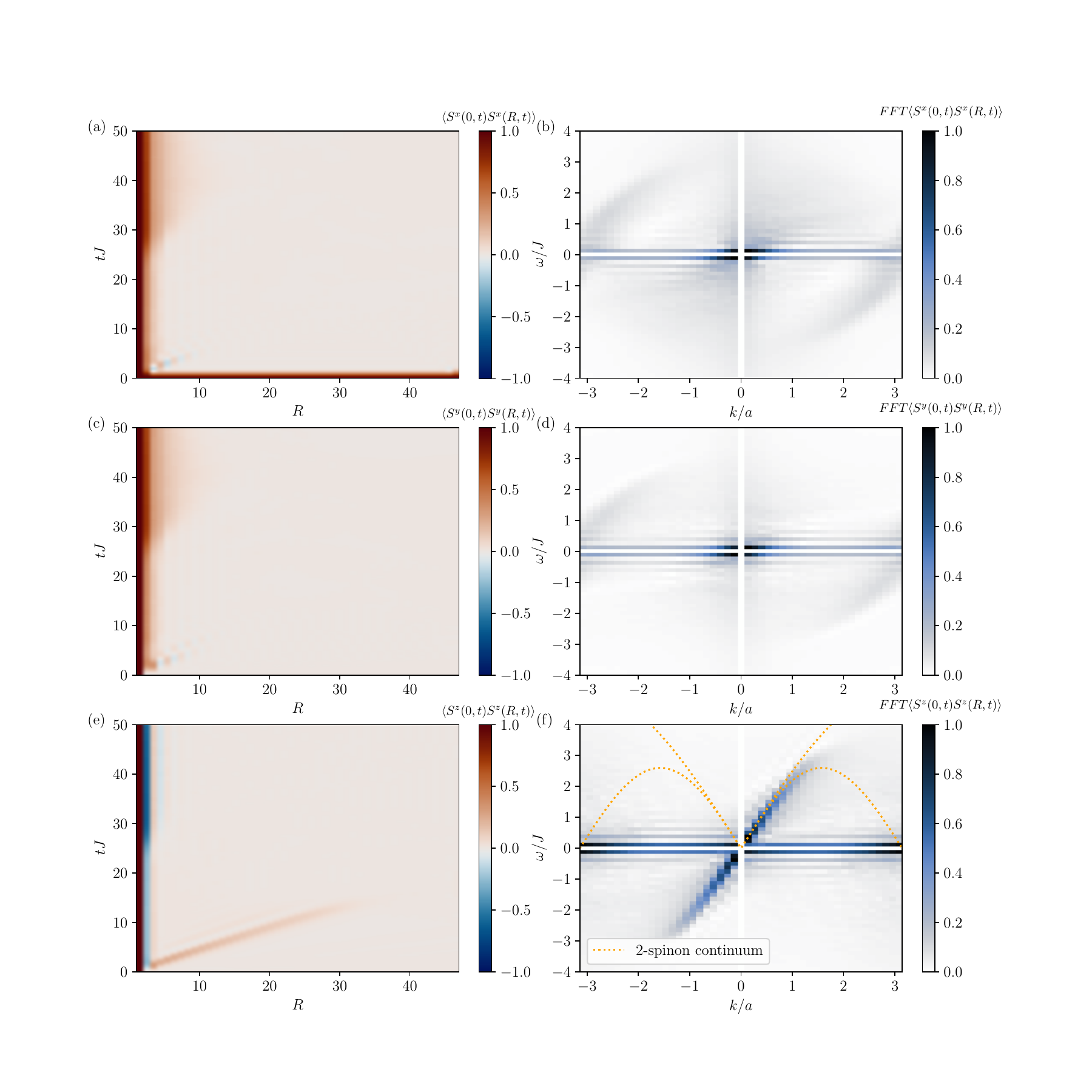}
    \caption{Global quench spectroscopy for $\Delta=-0.5$, again starting from an $x$-polarised product state as in Fig.~\ref{fig.XY50_GLOBAL} and showing correlation functions rather than local observables. As in Fig.~\ref{fig.XY-50}, the order in the XY plane rapidly melts, and only the observables in the $z$-direction give rise to a signal consistent with the spectrum.}
    \label{fig.XY-50_GLOBAL}
\end{figure}

\begin{figure}
    \centering
    \includegraphics[width=\linewidth]{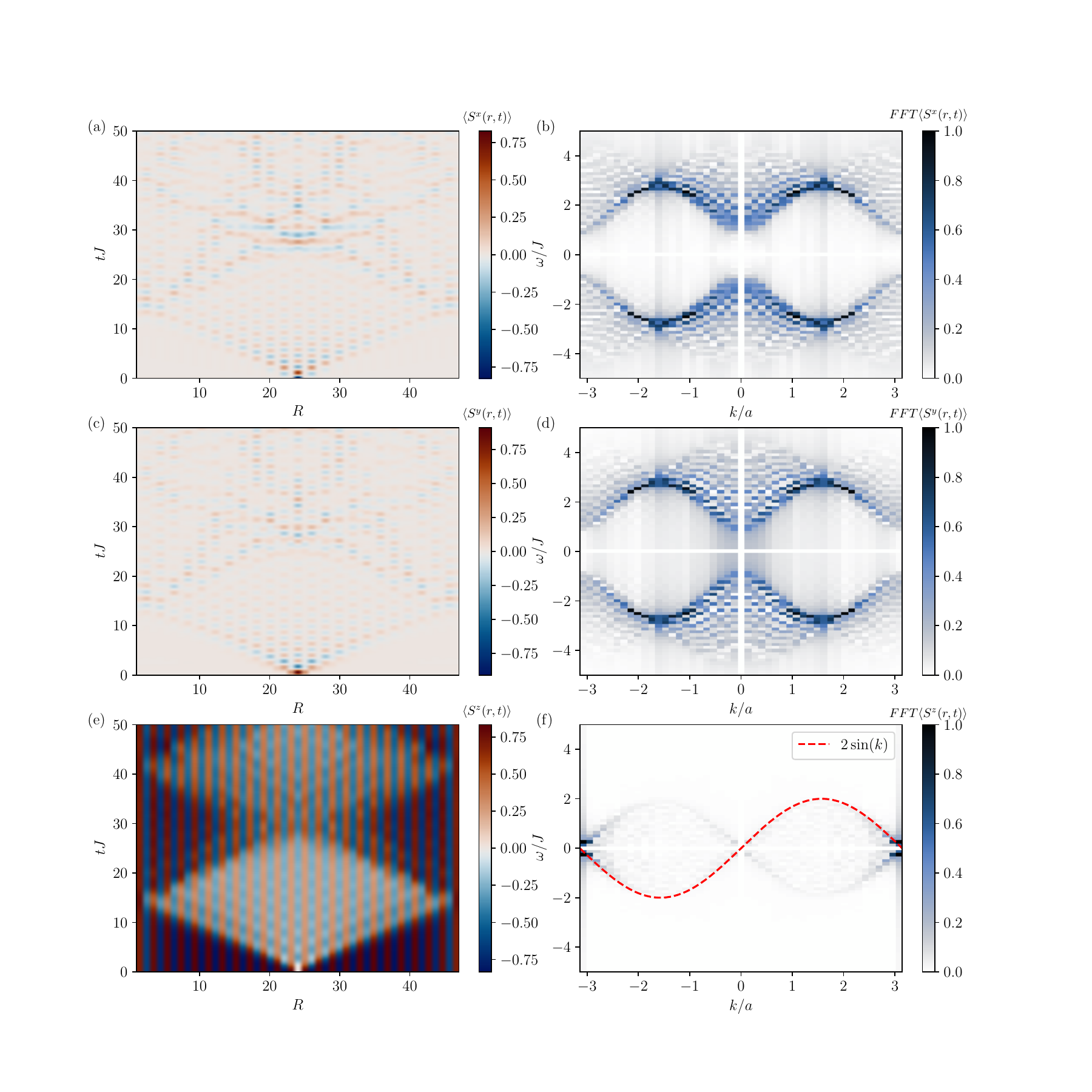}
    \caption{The same as Fig.~\ref{fig.FM3}, but for $\Delta=-2.5$, and now we start in the approximate ground state of the system (obtained using DMRG, with final bond dimension $\chi=14$), which is a $z$-aligned antiferromagnet. As this is the antiferromagnetic regime, we do not expect to be able to excite single excitations, but rather expect a gapped 2-spinon continuum. This is indeed what we see in the Fourier transforms of $\langle S^{x}(r,t) \rangle$ and $\langle S^{y}(r,t) \rangle$. In the Fourier transform of $\langle S^{z}(r,t) \rangle$, we again see a gapless signal similar to that seen in the other phases, but now with a modified periodicity due to the underlying 2-site unit cell in the AFM phase.}
    \label{fig.AFM25}
\end{figure}

\begin{figure}
    \centering
    \includegraphics[width=\linewidth]{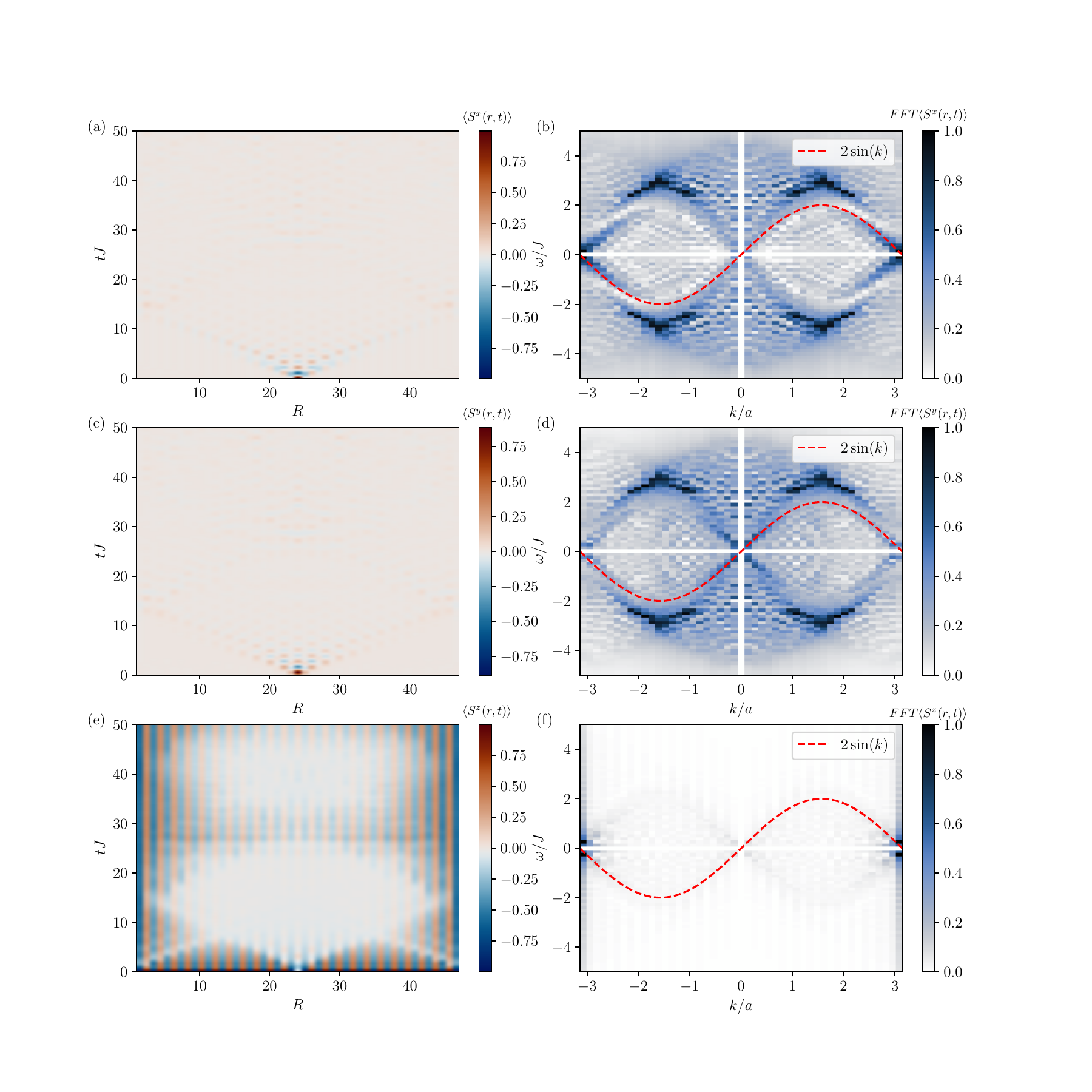}
    \caption{The same as Fig.~\ref{fig.AFM25}, but starting in the $\chi=1$ N\'eel state rather than a close approximation to the ground state. The two-spinon continuum is not only visible, but is enhanced over the case where the initial state is the ground state, likely due to the additional higher-energy excitations present in the initial state that help to `fill out' the upper portion of the continuum. The magnitude of the signal in the XY plane seems to decay more quickly, however, potentially making this more challenging to measure on noisy quantum computers. There is also an additional signal in the Fourier transforms of both $\langle S^{x}(r,t) \rangle$ and $\langle S^{y}(r,t) \rangle$ that appears just below the two-spinon continuum and matches the $2 \sin(k)$ algebraic divergence, again confirming that there are multiple types of excitation present in this initial state that is further from the ground state.}
    \label{fig.AFM25_NEEL}
\end{figure}

\begin{figure}
    \centering
    \includegraphics[width=\linewidth]{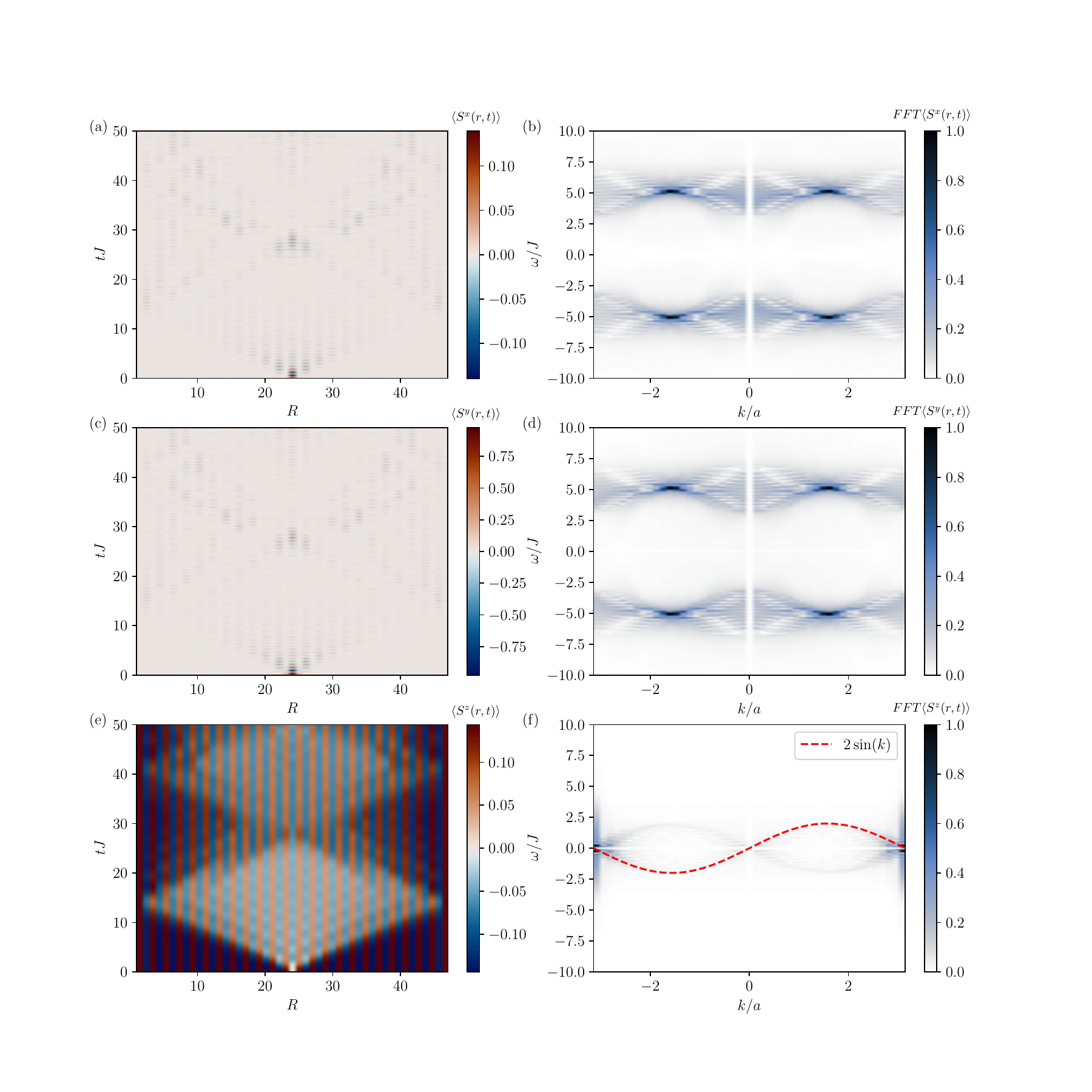}
    \caption{The same as Fig.~\ref{fig.AFM25}, but for $\Delta=-5$, again starting in the ground state (obtained with DMRG). As in Fig.~\ref{fig.AFM25}, the Fourier transforms of both $\langle S^{x}(r,t) \rangle$ and $\langle S^{y}(r,t) \rangle$ pick out the expected spectrum, while the Fourier transform of $\langle S^{z}(r,t) \rangle$ displays the $2 \sin(k)$ algebraic divergence.}
    \label{fig.AFM50}
\end{figure}

\begin{figure}
    \centering
    \includegraphics[width=\linewidth]{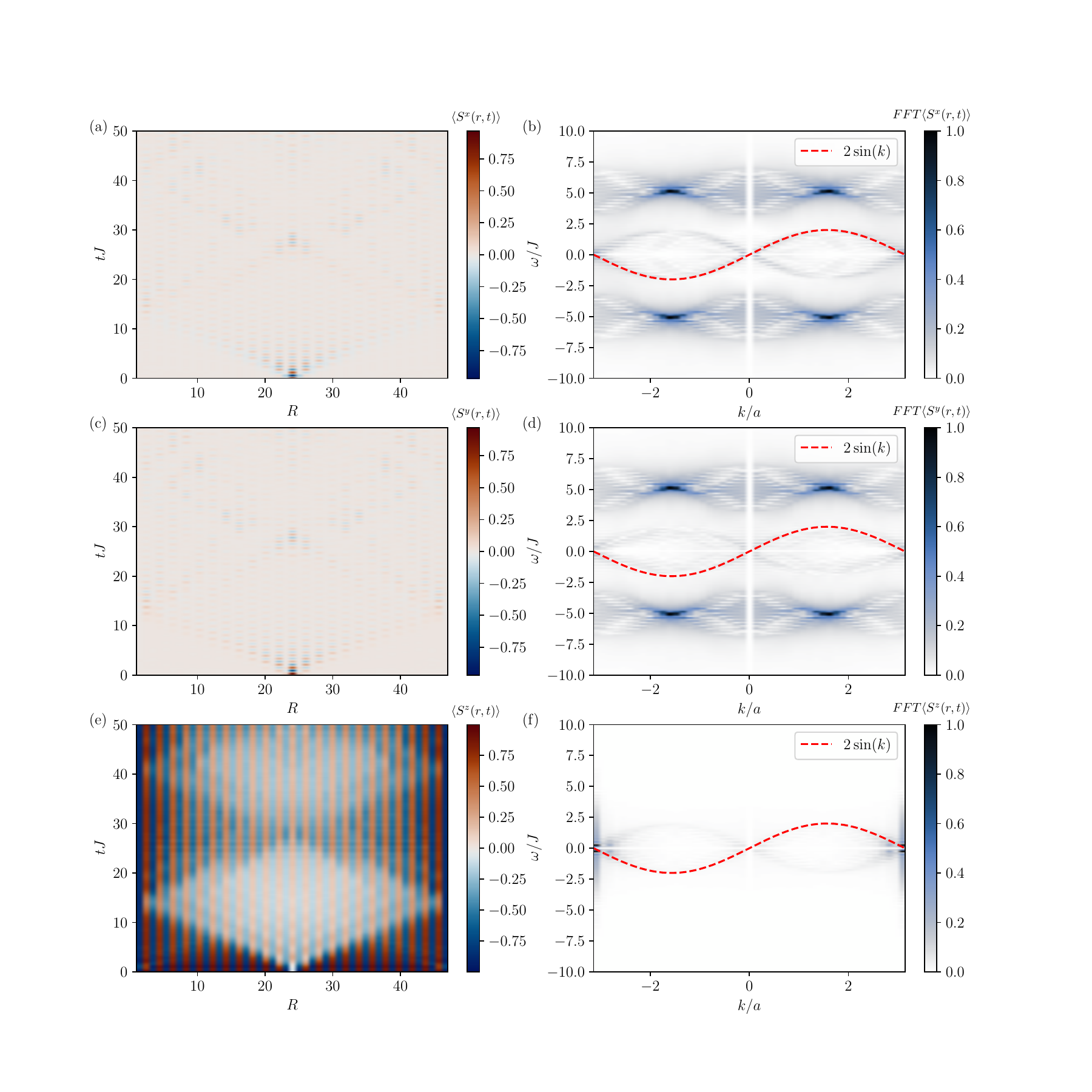}
    \caption{The same as Fig.~\ref{fig.AFM50}, but now starting in the $\chi=1$ N\'eel state. As in Figs~\ref{fig.AFM25} and ~\ref{fig.AFM25_NEEL}, the Fourier transforms of all variables display a signal that matches the $2 \sin(k)$ algebraic divergence.}
    \label{fig.AFM50_NEEL}
\end{figure}

Here, we give extensive classical benchmarks for the XXZ chain across all parameter regimes examined in the main text, and we show results for additional observables in order to illustrate the interplay of the quench and the probe. Figures~\ref{fig.FM1}, \ref{fig.FM2} and \ref{fig.FM3} show classical results in the FM phase. Figures~\ref{fig.XY99} and \ref{fig.XY99_GROUND} show the difference between starting in the ground state of the XY phase and the $x$-polarised product state. For the same parameters, we show how adding anistropy opens a gap in the observed spectrum (Fig.~\ref{fig.XY99_ANISOTROPIC}), consistent with the expected behaviour of the elementary excitation spectrum. Further into the XY phase, in Figure~\ref{fig.XY50} we give classical results in the regime where linear spin wave theory begins to break down. We also show the result of \emph{global quench spectroscopy} for the same parameters in Fig.~\ref{fig.XY50_GLOBAL}, demonstrating the effect of starting in the $x$-polarised product state without applying the local quench. Finally, Figures~\ref{fig.AFM25} and \ref{fig.AFM50} show results in the antiferromagnetic phase, starting in the true ground state, while Figures~\ref{fig.AFM25_NEEL} and \ref{fig.AFM50_NEEL} show the same parameter values, but with the N\'eel state as the initial state.

These benchmarks were performed using a matrix product state framework, making use of the ITensor library~\cite{Fishman+22}, and used the Time-Dependent Variational Principle algorithm to compute the non-equilibrium dynamics. We use system size $L=47$, timestep $dt = 0.05$, a maximum bond dimension of $\chi=256$, and an SVD cutoff of $\epsilon = 10^{-8}$. These results are intended to demonstrate the expected signals in each observable across each of the different parameter regimes, and as such we opted for a modest system size and reasonable level of accuracy in order to broadly survey a wide range of different parameters.

\clearpage
\bibliography{refs}